\documentclass[prd,aps,showpacs,epsf,floats,onecolumn]{revtex4}%
\usepackage{amssymb}
\usepackage{amsfonts}
\usepackage{amsmath}
\usepackage{graphicx}%
\providecommand{\U}[1]{\protect\rule{.1in}{.1in}}

\begin{document}
\title{\textbf{INFORMATION GEOMETRIC METHODS FOR COMPLEXITY}}
\author{\textbf{Domenico Felice}$^{1}$, \textbf{Carlo Cafaro}$^{2}$, \textbf{Stefano
Mancini}$^{3,4}$}
\affiliation{$^{1}$Max Planck Institute for Mathematics in the Sciences, Inselstrasse
\ 22--04103 Leipzig,Germany}
\affiliation{$^{2}$SUNY Polytechnic Institute, 12203 Albany, New York, USA}
\affiliation{$^{3}$School of Science and Technology, University of Camerino, 62032Camerino, Italy}
\affiliation{$^{4}$INFN-Sezione Perugia, Via A. Pascoli, I-06123 Perugia, Italy}

\begin{abstract}
Research on the use of information geometry (IG) in modern physics has
witnessed significant advances recently. In this review article, we report on
the utilization of IG methods to define measures of complexity in both
classical and, whenever available, quantum physical settings. A paradigmatic
example of a dramatic change in complexity is given by phase transitions
(PTs). Hence we review both global and local aspects of PTs described in terms
of the scalar curvature of the parameter manifold and the components of the
metric tensor, respectively. We also report on the behavior of geodesic paths
on the parameter manifold used to gain insight into the dynamics of PTs. Going
further, we survey measures of complexity arising in the geometric framework.
In particular, we quantify complexity of networks in terms of the Riemannian
volume of the parameter space of a statistical manifold associated with a
given network. We are also concerned with complexity measures that account for
the interactions of a given number of parts of a system that cannot be
described in terms of a smaller number of parts of the system. Finally, we
investigate complexity measures of entropic motion on curved statistical
manifolds that arise from a probabilistic description of physical systems in
the presence of limited information. The Kullback-Leibler divergence, the
distance to an exponential family and volumes of curved parameter manifolds,
are examples of essential IG notions exploited in our discussion of
complexity. We conclude by discussing strengths, limits, and possible future
applications of IG methods to the physics of complexity.

\end{abstract}

\maketitle

\begin{quotation}
\bigskip\pagebreak

\textbf{LEAD PARAGRAPH}. Describing and, to a certain extent, understanding
the concept of complexity have been investigated in a variety of research
fields. Various ad hoc formal definitions and interpretations of the concept
of complexity of physical systems have been proposed depending on the specific
domain of interest \cite{adami02}. Furthermore, since the science of
complexity is growing rapidly, new complexity measures are continuously being
developed. For instance, quantifying the degree of organization of a physical
system often resorts to some type of entropy function stemming from the
Shannon information entropy which can be shown to be itself equivalent to
standard forms of entropy that appear in physics \cite{peres95}. In
particular, Information Geometry (IG) provides relevant tools for defining a
geometric entropy for either measuring complex networks or identifying the
most relevant interacting subsystems within composite complex systems at
different scales of observation. Moreover, when combined with statistical
inference tools, IG offers a convenient platform to characterize the
complexity of statistical predictions in the presence of partial knowledge of
the system being considered.

In this manuscript, starting from the observation that a change in the
complexity behavior of a physical system can be geometrically detected in the
proximity of a phase transition, we review information geometric aspects of
both phase transitions and complexity measures in classical and quantum settings.

Given the wide range of applicability of methods of information geometry to
the science of complexity, we argue that the collection of findings selected
and discussed in our work may be useful to motivate other scientists to find a
possible unifying information geometric complexity measure of universal applicability.

\pagebreak
\end{quotation}

\section{Introduction}

Information geometry (IG) is the application of methods of modern geometry to
sets of probability distributions (statistical models). IG has arisen from
investigations of invariant geometrical structure within the framework of
statistical inference. It provides a Riemannian metric and two dual affine
connections in a statistical model \cite{amari85}. Beyond the statistical
inference, the geometric structure of the space of probability distributions
plays a fundamental role in wider areas of information sciences, such as
machine learning, signal processing, optimization as well as statistical
physics and mathematical theory underlying neural networks \cite{amari16}.
Furthermore, the IG of statistical models can also be extended to quantum
mechanics (QM), with QM regarded as a non-commutative extension of probability
theory. However, the study of the so-called quantum information geometry has
started only very recently and it can be essentially said to be in its infancy
stage \cite{amari}. IG therefore has become a framework of great interest due
to the insight it provides into information sciences and wherever information
theoretical quantities like relative entropies find a natural geometric
interpretation \cite{Ay17}. As such information geometric measures can be
relevant in the science of complexity.

Phase transitions are the most impressive examples in nature of emergent
phenomena. They are fairly well-understood from a theoretical standpoint and
characterized by a sharp change in the complexity of the physical system that
exhibits such emergent phenomena when a suitable control parameter exceeds a
certain critical value. In classical thermodynamics \cite{beretta05, huang87},
a first-order phase transition is a phase transition of a pure substance if it
is characterized by two phases which, when coexisting, exhibit different
values of the first-order derivatives of the chemical potential with respect
to temperature and pressure. The first-order derivatives of the chemical
potential can be expressed in terms of the specific volume and the specific
entropy. On the contrary, a second-order phase transition is a feature of pure
substances that admit pairs of phases which, when coexisting, do not differ in
the values of the specific volume and the specific entropy. Instead, they
differ in crystallographic configuration, magnetic moment configuration (that
is, ferromagnetic properties), symmetry, or other features, including the
values of the second-order derivatives of the chemical potential with respect
to temperature and pressure. From a quantum-mechanical perspective,
first-order quantum phase transitions (QPTs, \cite{sachdev99}) correspond to
discontinuities in the first derivative of the ground state energy density.
For instance, in many-body quantum systems, examples of first-order QPTs are
points of degeneracy of the Hamiltonians \cite{hamma06}. Due to their limited
physical interest, we do not take into consideration higher-order QPTs that
correspond to discontinuities in the higher order derivatives of the ground
state energy density.

Phase transitions were originally investigated in the framework of the
Riemannian geometrization of classical Hamiltonian dynamics in Ref.
\cite{caiani97}. Two main questions were addressed there: first, are there
peculiar emerging features in the geometric properties associated with the
chaotic Hamiltonian dynamics of a system in the proximity of a phase
transition? For example, a suitable geometric indicator of complexity could be
represented by the average curvature properties of the curved manifold
underlying the dynamics. Second, how do curvature fluctuations and Lyapunov
exponents behave at critical points? These questions were studied for the
planar Heisenberg $XY$-model \cite{caiani97}. The cuspid-like behavior of the
largest Lyapunov exponent as a function of the temperature (defined as the
time average of the kinetic energy per degree of freedom) at the critical
temperature was reported based on both numerical and analytical arguments. The
same cuspid-like behavior was reported for the fluctuations of the Gaussian
curvature at the critical temperature. Based on these findings, the so-called
topological conjecture was advanced: in the vicinity of a second-order phase
transition, the manifold underlying the chaotic Hamiltonian dynamics undergoes
a topology change. Such a topological conjecture found further numerical
support in Ref. \cite{franzosi99} by studying the numerically-computed
behavior of the variance of the scalar curvature of the manifold underlying a
classical $\varphi^{4}$ lattice model. Furthermore, the conjecture was later
analytically supported by the investigation of a mean-field model in Ref.
\cite{firpo98}. For a theoretical discussion on the topological origin of a
phase transition in a mean-field model, we refer to Ref. \cite{casetti99}.
Within the IG\ framework, phase transitions are studied by analyzing the
scalar curvature properties and the behavior of geodesics on the curved
parameter manifold associated with the physical system being considered once a
suitable metric tensor is introduced. For classical physical systems, the
parameters that characterize the metric tensor are generally described by
thermodynamical control parameters such as pressure, volume, and temperature.
For quantum physical systems, instead, such parameters are in general the
coupling constants that appear in the Hamiltonian of the system. The first
investigations of the geometric structure of equilibrium thermodynamics were
performed by Weinhold and reported in Refs.
\cite{weinhold75,weinhold75a,weinhold75b,weinhold75c,weinhold76}. However, the
origin of the application of Riemannian geometric methods into phase
transition investigations is due to the inclusion of Einstein's fluctuation
theory \cite{einstein04,einstein10} in the axioms of thermodynamics. Indeed,
in this case, it happens that thermodynamic systems can be described in terms
of Riemannian manifolds \cite{ruppeiner79}. In particular, within this
geometric framework, the thermodynamic curvature is a suitable indicator of
the nature (attractive or repulsive) of the interparticle interactions in the
thermodynamic system. The original finding that lead to such a physical
interpretation of the thermodynamical curvature was the fact that in an ideal
gas there is no effective interparticle interaction and, remarkably, the
curvature of the manifold that describes a classical ideal gas is identically
zero \cite{ruppeiner79}. This original interpretation of curvature was due to
Ruppeiner, according to whom the curvature is positive when repulsive
interactions are dominant, and negative when attractive interactions dominate.
It is worthwhile mentioning that there exist alternative interpretations of
the sign of the thermodynamic curvature. For instance, according to Janyszek
and Mrugala \cite{jany89}, the scalar curvature of the parameter manifold
measures the stability of the physical system being considered. In particular,
the larger the curvature, the more stable the system. The scientific finding
that motivated such an interpretation was the fact that for a one-dimensional
Ising model with short-range interactions, one observes large fluctuations
(large curvature) at low temperatures and zero magnetic field in the
ferromagnetic case and small fluctuations (small curvature) in the
antiferromagnetic case. More specifically, it was reported in\ Ref.
\cite{jany89} that since the scalar curvature of the parameter space
approaches infinity in the vicinity of critical points, the inverse of the
scalar curvature was interpreted as a measure of the stability of the magnetic
system being analyzed.

A major issue in complex systems, besides recognizing emergent phenomena like
phase transitions, is to quantify their degree of complexity. In fact the
concepts of information and distance play a key role in defining the notion of
complexity of physical systems even though its origin is not completely
understood \cite{L88, GM95, Lopez95, F98,W84, W85, R95}. According to the
domain of interest, different definitions of complexity and methods of
measuring it have been proposed and are continuously being proposed since the
science of complexity is still rapidly developing \cite{GM95, L88}. Within the
framework of classical physics, measures of complexity are understood in a
more satisfactory manner. For instance, the Kolmogorov-Sinai metric entropy
\cite{K65, K68}, the sum of all positive Lyapunov exponents \cite{P77}, is a
powerful indicator of unpredictability in classical dynamical systems. It is a
measure of the so-called algorithmic complexity of classical trajectories
\cite{B83, B00, S89, W78}. Other known measures of complexity include, but are
not limited to, the logical depth \cite{B90}, the thermodynamic depth
\cite{S88}, the computational complexity \cite{P94} and stochastic complexity
\cite{R86}. Roughly speaking the logical and thermodynamic depths identify as
complex whatever can be reached only through a difficult path. Each one of
these complexity indicators captures, to a certain extent, our intuitive ideas
about the meaning of complexity. Some of them just apply to computational
tasks and, unfortunately, only very few of them may be generalized so that
their field of applicability can include the investigation of actual physical
processes. From an ideal viewpoint, a good definition of complexity should be
both mathematically rigorous and have a clear intuitive interpretation so as
to allow for the study of complexity-related problems in both computation
theory and statistical physics. Naturally, a quantitative measure of
complexity is truly useful if its range of applicability is not limited to a
few unrealistic applications. Therefore, in order to properly define measures
of complexity, not only should the motivations for defining such a measure be
clearly stated, but also what feature the measure is intended to capture
should be neatly described. Quantifying the complexity of complex networks
where both the concepts of probability distributions and structure play
leading roles \cite{franzosi16}, studying the complexity of structures that
characterize ensemble of objects \cite{olbrich08}, and, finally, investigating
the complexity of statistical models \cite{ali17} have witnessed an increasing
research interest in recent years.

To the best of our knowledge, the first preliminary investigations on the use
of methods of information geometry to quantify complexity appeared in 2006 in
Refs. \cite{ay06, Cafaro06}. In Ref. \cite{ay06}, Ay and collaborators
exploited the information geometric framework of hierarchies of exponential
families in order to propose a theory of complexity measures for finite random
fields. In Ref. \cite{Cafaro06}, Cafaro and collaborators investigated the
complexity of geodesic paths on negatively curved statistical manifolds
underlying the entropic dynamics of a two-dimensional system in the presence
of incomplete knowledge on its microscopic degrees of freedom.

In this paper, we survey the application of IG methods to characterize aspects
of complexity in applied science.\textbf{ }By using IG techniques, we report
on properties of both classical and quantum PTs in physical systems that
exhibit critical behavior. Global and local aspects of PTs specified by means
of the scalar curvature of the parameter manifold and the components of the
metric tensor, respectively, will be reviewed. We shall also discuss the
behavior of geodesic paths on the parameter manifold used to gain insights
into the dynamics of PTs. We will then go through a geometric entropy that
arises by associating smooth systems (Riemannian manifolds) to discrete
systems (networks). In particular, we will show that such a geometric entropy
is capable of detecting the Erd\"{o}s-R\'{e}nyi phase transition in uniform
random graphs. Therefore, such geometric entropy will also be discussed as a
candidate for measuring networks complexity \cite{franzosi16}. Following this
idea of characterizing complexity in terms of IG methods, we shall report on
complexity measures displaying interactions of a fixed number of parts of a
composite system that cannot be described by means of a smaller number of
parts of the system \cite{olbrich08}. Finally we shall review complexity
measures of entropic motion on curved statistical manifolds that originate
from a probabilistic description of physical systems in the presence of
partial knowledge \cite{ali17}.

The layout of the manuscript is as follows. In Section II, we present the
essential tools of information geometry needed to characterize the concepts of
complexity and phase transitions in both classical and quantum physical
systems. In Section III, we discuss the use of IG to characterize classical
phase transitions of second-order that occur in a variety of physical
settings, including the one-dimensional Ising model, the one-dimensional
$q$-state Potts model, and the Ising model on dynamical planar random graphs.
Methods of IG are used in Section IV to single out the Erd\"{o}s-R\'{e}nyi
phase transition in uniform random graphs and subsequently to characterize the
complexity of networks. Further methods of IG are presented in Section V to
study the complexity of systems with multiple interacting units at different
scales of interest, and, finally, to quantify the complexity of macroscopic
predictions in the presence of limited information in Section VI. In Section
VII we review the use of IG techniques to study quantum phase transitions of
second-order in the periodic XY spin chain in a transverse magnetic field and
in the Dicke model of quantum optics. In particular, for the former model, we
discuss both a curvature-based and a geodesic-based analysis. Finally, after
briefly addressing the notion of complexity of quantum evolution, we discuss
both an information geometric perspective on the phenomena of softening of
chaos by quantization and a statistical complexity measure of probability
distributions that characterize quantum mechanical systems exhibiting
nontrivial dynamics. A summary of conclusive results together with several
considerations of unresolved issues concerning the use of information
geometric methods to quantify the complexity of quantum mechanical systems are
reported in Section VIII. Useful tools of differential geometry appear in
Appendix A.


\section{Basics of information geometry}

In this section, referring to \cite{amari85,amari,amari16}, we introduce the
basic notions of IG, most of which will be employed in the following parts of
the manuscript.

A statistical manifold $\mathcal{M}$ can be described as a set of parametrized
probability distribution functions {(pdfs)}, with a tensor metric $g$ and an
affine connection $\nabla$. Let $p\left(  \boldsymbol{x};\boldsymbol{\theta
}\right)  $ denote the pdf of $\boldsymbol{x}$ specified by a parameter vector
$\boldsymbol{\theta}=(\theta^{1},\ldots,\theta^{n})$. We assume the regularity
conditions that the density functions $p(\boldsymbol{x};\boldsymbol{\theta})$
exist on some carrier measure of the space $\mathcal{X}$ of $\boldsymbol{x}$,
that
\begin{equation}
p(\boldsymbol{x};\boldsymbol{\theta})>0\text{,}\ \text{and}\int_{\mathcal{X}%
}\ p(\boldsymbol{x};\boldsymbol{\theta})d\boldsymbol{x}=1
\end{equation}
for all $\boldsymbol{x}$ in the common domain $\mathcal{X}$. Furthermore, the
mapping $\varphi:p(\boldsymbol{x};\boldsymbol{\theta})\mapsto
\boldsymbol{\theta}$ is an homeomorphism between $\mathcal{M}$ and the open
set $\Theta\subset\mathbb{R}^{n}$ of the parameters $\boldsymbol{\theta}$. In
addition, we assume that $(\mathcal{M},\varphi)$ is a differentiable
structure; in such a way\textbf{,} $\mathcal{M}$ is an $n$-dimensional
$C^{\infty}$-manifold. In addition, let
\begin{equation}
l(\boldsymbol{x};\boldsymbol{\theta})\overset{\text{def}}{=}\ln
p(\boldsymbol{x};\boldsymbol{\theta}).
\end{equation}
Then, we may regard every point $\boldsymbol{\theta}$ of $\Theta$ as carrying
a function $l(\boldsymbol{x};\boldsymbol{\theta})$ of $\boldsymbol{x}$. In
this way, we can have a useful representation of\textbf{ }vector fields in
terms of the basis $\{\partial_{i}l(\boldsymbol{x};\boldsymbol{\theta})\}$
with $\partial_{i}\overset{\text{def}}{=}\frac{\partial}{\partial\theta^{i}}$.
This representation has the advantage that a vector field has zero expectation
value with respect to the pdf $p(\boldsymbol{x};\boldsymbol{\theta})$,
\begin{equation}
\mathbb{E}_{p}\left[  \partial_{i}l(\boldsymbol{x},\boldsymbol{\theta
})\right]  \overset{\text{def}}{{=}}\underset{\mathcal{X}}{\int}p\left(
\boldsymbol{x};\boldsymbol{\theta}\right)  \frac{\partial\ln p\left(
\boldsymbol{x};\boldsymbol{\theta}\right)  }{\partial\theta^{i}}%
d\boldsymbol{x}=\frac{\partial}{\partial\theta^{i}}\underset{\mathcal{X}}%
{\int}p\left(  \boldsymbol{x};\boldsymbol{\theta}\right)  d\boldsymbol{x}%
=0\text{.} \label{ev1}%
\end{equation}
In addition to the connection $\nabla$, the statistical manifold $\mathcal{M}$
also carries a dual connection $\nabla^{\ast}$ which is defined as
\cite{amari}
\begin{equation}
g(\nabla_{X}^{\ast}Y,Z)\overset{\text{def}}{=}Xg(Y,Z)-g(\nabla_{X}Y,Z)\text{,}
\label{dual}%
\end{equation}
with $X$, $Y$, and $Z$ denoting vector fields on $\mathcal{M}$. A typical
example of pdfs is provided by the so-called exponential family \cite{amari82}%
. For such a family, the notion of distinguishability between pdfs is
quantified by means of the so-called Fisher--Rao information metric tensor
\cite{amari82}
\begin{equation}
g_{ij}\left(  \boldsymbol{\theta}\right)  \overset{\text{def}}{=}%
\mathbb{E}_{p}\left[  \partial_{i}l(\boldsymbol{x};\boldsymbol{\theta
})\partial_{j}l(\boldsymbol{x};\boldsymbol{\theta})\right]  \text{.}
\label{F-R}%
\end{equation}
Furthermore, the affine connection $\nabla$ related to $g_{ij}$ in Eq.
(\ref{F-R}) is known as the exponential connection and can be defined in terms
of the following Christoffel coefficients
\begin{equation}
\Gamma_{ijk}\overset{\text{def}}{=}g\left(  \nabla_{\partial_{i}}\partial
_{j},\partial_{k}\right)  =\mathbb{E}_{p}\left[  \partial_{i}\partial
_{j}l(\boldsymbol{x},\boldsymbol{\theta})\partial_{k}l(\boldsymbol{x}%
,\boldsymbol{\theta})\right]  \text{.} \label{expcon}%
\end{equation}
Associated to the exponential connection $\nabla$, the dual connection
$\nabla^{\ast}$ is given by
\begin{equation}
\Gamma_{ijk}^{\ast}\overset{\text{def}}{=}g\left(  \nabla_{\partial_{i}}%
^{\ast}\partial_{j},\partial_{k}\right)  =\mathbb{E}_{p}\left[  \left(
\partial_{i}\partial_{j}l(\boldsymbol{x},\boldsymbol{\theta})+\partial
_{i}l(\boldsymbol{x},\boldsymbol{\theta})\partial_{j}l(\boldsymbol{x}%
,\boldsymbol{\theta})\right)  \partial_{k}l(\boldsymbol{x},\boldsymbol{\theta
})\right]  \text{.} \label{expcon*}%
\end{equation}
Indeed, a statistical structure on the manifold $\mathcal{M}$ is given in
terms of a dualistic structure $(g,\nabla,\nabla^{\ast})$. When an exponential
family is considered, it is well-known that the Riemann curvature tensor
$\mathcal{R}$ with respect to the exponential affine connection $\nabla$
{[Appendix \ref{DiffGeom}]} is zero implying that such a statistical manifold
is $\nabla$-flat. In this particular case, even the Riemann curvature tensor
$\mathcal{R}^{\ast}$ with respect to the dual affine connection $\nabla^{\ast
}$ equals zero. Hence, it is evident that the exponential family is a dually
flat statistical manifold \cite{amari}. Given a statistical geometric
structure $(\mathcal{M},g,\nabla,\nabla^{\ast})$, the information about the
statistical geometric structure is retained in a two points differentiable
real-valued function $\mathcal{D}$ \cite{matsumoto93}. This is a divergence
function $\mathcal{D}:\mathcal{M}\times\mathcal{M}\rightarrow\mathbb{R}^{+}$
for the manifold $\mathcal{M}$ which is a distance-like measure of the
separation between two points $p,q\in\mathcal{M}$ such that
\begin{equation}
\mathcal{D}[p,q]\geq0\text{, and }\mathcal{D}[p,q]=0\Leftrightarrow p=q.
\label{divergence}%
\end{equation}
Given a coordinate system $[\theta^{i}]$, one can represent a pair of points
$(p,q)\in\mathcal{M}\times\mathcal{M}$ as a pair of coordinates $([\theta
^{i}(p)],[\theta^{i}(q)])$. In particular, the metric tensor $g$ and the
connection $\nabla$ provide a third order approximation for $\mathcal{D}$
\cite{amari},
\begin{align}
\mathcal{D}[p,q]  &  =\frac{1}{2}g_{ij}\Delta\theta^{i}\Delta\theta^{j}%
+\frac{1}{6}h_{ijk}\Delta\theta^{i}\Delta\theta^{j}\Delta\theta^{k}%
+o(\Vert\Delta\theta\Vert^{3})\label{Thrid}\\
&  =\frac{1}{2}g_{ij}\Delta\theta^{i}\Delta\theta^{j}-\frac{1}{6}h_{ijk}%
^{\ast}\Delta\theta^{i}\Delta\theta^{j}\Delta\theta^{k}+o(\Vert\Delta
\theta\Vert^{3}),\nonumber
\end{align}
where $\Delta\theta^{i}\overset{\text{def}}{=}\theta^{i}(p)-\theta^{i}(q)$ and
$o(\Vert\Delta\theta\Vert^{3})$ is a term that vanishes at a rate faster than
$\Vert\Delta\theta\Vert^{3}$ as $p$ tends to $q$. The coefficients $h_{ijk}$
and $h_{ijk}^{\ast}$ define a $3$-covariant symmetric tensor which, in terms
of the metric tensor $g$ and the connection coefficients $\Gamma_{ijk}$ of the
affine connection $\nabla$, is given by
\begin{equation}
h_{ijk}=\partial_{i}g_{jk}+\Gamma_{jk,i}\text{, and, }h_{ijk}^{\ast}%
=\partial_{i}g_{jk}+\Gamma_{jk,i}^{\ast}\text{,} \label{symmetrictensor}%
\end{equation}
respectively, where $\partial_{i}\overset{\text{def}}{=}\frac{\partial
}{\partial\theta^{i}}$. Conversely, given a divergence function $\mathcal{D}$
for $(\mathcal{M},g,\nabla,\nabla^{\ast})$ and the standard coordinate systems
$([\theta^{i}],[\theta^{\prime i}])$, the statistical geometric structure is
recovered as follows \cite{amari},
\begin{equation}
g_{ij}=-\partial_{i}\partial_{j}^{\prime}\mathcal{D}\left[  p,q\right]
_{|p=q}\text{, and }\Gamma_{ij,k}=-\partial_{i}\partial_{j}\partial
_{k}^{\prime}\mathcal{D}\left[  p,q\right]  _{|p=q}\text{,}
\label{geomfromdiv}%
\end{equation}
where $\partial_{i}^{\prime}\overset{\text{def}}{=}\frac{\partial}%
{\partial\theta^{\prime i}}$ is the partial derivative with respect to the
second variable. Despite the infinitely many ways in which one can define a
divergence function that generates a given geometry, it is often useful to
define a divergence function which is, in some sense, canonical. A canonical
divergence satisfies desirable properties such as the generalized Pythagorean
theorem and the geodesic projection theorem \cite{amari}. In the particular
case of a dually flat statistical manifold, a canonical divergence (a
Bregman-type of divergence) was introduced by Amari and Nagaoka in Ref.
\cite{amari}. An exponential model is a statistical manifold that consists of
probability distribution functions $e(\boldsymbol{x};\boldsymbol{\theta})$
defined as
\begin{equation}
e(\boldsymbol{x};\boldsymbol{\theta})\overset{\text{def}}{=}\exp\left[
C(\boldsymbol{x})+\sum_{i=1}^{n}\theta^{i}F_{i}(\boldsymbol{x})-\psi
(\boldsymbol{\theta})\right]  \text{,} \label{exp}%
\end{equation}
where $\{C,F_{1},\ldots,F_{n}\}$ are functions on the sample space
$\mathcal{X}$ and $\psi$ is a function on the space $\Theta$ of the parameters
$\boldsymbol{\theta}$. From the normalization condition $\int_{\mathcal{X}%
}e(\boldsymbol{x};\boldsymbol{\theta})d\boldsymbol{x}=1$, we obtain
\begin{equation}
\psi(\boldsymbol{\theta})=\ln\left\{  \int_{\mathcal{X}}\exp\left[
C(\boldsymbol{x})+\sum_{i=1}^{n}\theta^{i}F_{i}(\boldsymbol{x})-\psi
(\boldsymbol{\theta})\right]  \ d\boldsymbol{x}\right\}  \text{.}
\label{potential}%
\end{equation}
We refer to this function $\psi$ in Eq. (\ref{potential}) as the potential
function of the exponential family. Indeed, the family $\mathbf{E}%
=\{e(\boldsymbol{x};\boldsymbol{\theta})\}$ is a statistical manifold
with\textbf{ }the metric tensor $g$ as in Eq. \eqref{F-R} and the connections
$\nabla$ and $\nabla^{\ast}$ as in Eqs. \eqref{expcon},\eqref{expcon*},
respectively. {We shall refer} to such an information geometric structure as
$(\mathbf{E},g,\overset{e}{\nabla},\overset{e}{\nabla^{\ast}})$. From the
definition of an exponential family given in Eq. \eqref{exp}, we have
\begin{equation}
\partial_{i}l(\boldsymbol{x};\boldsymbol{\theta})=F_{i}(\boldsymbol{x}%
)-\partial_{i}\psi(\boldsymbol{\theta})\text{, and }\partial_{i}\partial
_{j}l(\boldsymbol{x};\boldsymbol{\theta})=-\partial_{i}\partial_{j}%
\psi(\boldsymbol{\theta})\text{.} \label{potentials}%
\end{equation}
Furthermore, note that from Eqs. \eqref{ev1} and \eqref{potentials}, it is
possible to rewrite the components of the Fisher-Rao information metric
$g_{ij}\left(  \boldsymbol{\theta}\right)  $ as follows,
\begin{equation}
g_{ij}\left(  \boldsymbol{\theta}\right)  =-\mathbb{E}_{p}[\partial
_{i}\partial_{j}l(\boldsymbol{x};\boldsymbol{\theta})]=\partial_{i}%
\partial_{j}\psi(\boldsymbol{\theta})\text{.} \label{geqref}%
\end{equation}
In addition, from\textbf{ }Eqs. \eqref{ev1} and \eqref{potentials}, we
determine that the connection coefficients of the affine connection
$\overset{e}{\nabla}$ are given by
\begin{equation}
\overset{e}{\Gamma}_{ij,k}=-\partial_{i}\partial_{j}\psi(\boldsymbol{\theta
})\mathbb{E}_{\boldsymbol{\theta}}[\partial_{k}l(\boldsymbol{x}%
;\boldsymbol{\theta})]=0\text{.} \label{expChrist}%
\end{equation}
Eq. (\ref{expChrist}) implies that the exponential family $\mathbf{E}$ is a
flat manifold with respect to the affine connection $\overset{e}{\nabla}$ and
$[\theta^{i}]$ is called an affine coordinate system of $\overset{e}{\nabla}$
for $\mathbf{E}$. Associated with $[\theta^{i}]$, a dual coordinate system
$[\eta_{i}]$ can be defined as $\eta_{i}\overset{\text{def}}{=}\frac
{\partial\psi}{\partial\theta^{i}}$. The coordinates $[\theta^{i}]$ and
$[\eta_{i}]$ are dual with respect to the Fisher-Rao information metric $g$ in
the sense that the following relation holds true,
\begin{equation}
g(\partial_{i},\partial^{j})=\delta_{i}^{j}, \label{dualcoord}%
\end{equation}
where $\partial^{i}\overset{\text{def}}{=}\frac{\partial}{\partial\eta_{j}}$.
It is straightforward from the definition of $\eta_{j}$ that Eq.
(\ref{geqref}) is equivalent to
\begin{equation}
g_{ij}=\frac{\partial\eta_{j}}{\partial\theta^{i}}\text{, and }g^{ij}%
=\frac{\partial\theta^{i}}{\partial\eta_{j}}\text{,} \label{equivdual}%
\end{equation}
where $g^{ij}$ are the components of the inverse of the Fisher-Rao information
metric $g$. Therefore, together with the $\overset{e}{\nabla}$-flatness, the
manifold $\mathbf{E}$ is also $\overset{e}{\nabla^{\ast}}$-flat and $[\eta
_{i}]$ becomes as an affine coordinate system of $\overset{e}{\nabla^{\ast}}$
for $\mathbf{E}$. For this reason, one says that the exponential manifold
$(\mathbf{E},g,\overset{e}{\nabla},\overset{e}{\nabla^{\ast}})$ endowed with
the Fisher-Rao information metric $g$ {and exponential connection $\overset
{e}{\nabla}$} is a dually flat statistical manifold. To complete the
investigation of dualistic geometric structure of $\mathbf{E}$, we can
consider the dual potential function $\varphi$ of $\psi$ being defined as,
\begin{equation}
\varphi(\boldsymbol{\theta})\overset{\text{def}}{=}\theta^{i}\eta_{i}%
-\psi(\boldsymbol{\theta})\text{.} \label{legendre}%
\end{equation}
It immediately follows from \eqref{legendre} that $\theta^{i}=\partial
^{i}\varphi$ and, as a consequence, from Eq. \eqref{equivdual} we find
\begin{equation}
g^{ij}=\partial^{i}\partial^{j}\varphi\text{.} \label{potentialfi}%
\end{equation}
At this point, given the dual flat manifold $(\mathbf{E},g,\overset{e}{\nabla
},\overset{e}{\nabla^{\ast}})$, the coordinate systems $[\theta^{i}]$ and
$[\eta_{i}]$, and the potentials $\psi$ and $\varphi$, the canonical
divergence $\mathcal{D}$ becomes
\begin{equation}
\mathcal{D}[p_{\boldsymbol{\theta}},p_{\boldsymbol{\theta}^{\prime}}%
]=\psi(p_{\boldsymbol{\theta}})-\varphi(p_{\boldsymbol{\theta}^{\prime}%
})-\theta^{i}\eta_{i}\text{.} \label{candiv}%
\end{equation}
In the case of an exponential family, $\mathcal{D}[p_{\boldsymbol{\theta}%
},p_{\boldsymbol{\theta}^{\prime}}]$ in Eq. (\ref{candiv}) becomes the
well-known Kullback--Leibler divergence,
\begin{equation}
D_{\mbox{\tiny KL}}[p_{\boldsymbol{\theta}},p_{\boldsymbol{\theta}^{\prime}%
}]\overset{\text{def}}{=}\int_{\mathcal{X}}p(\boldsymbol{x};\boldsymbol{\theta
}^{\prime})\ln\left(  \frac{p(\boldsymbol{x};\boldsymbol{\theta}^{\prime}%
)}{p(\boldsymbol{x};\boldsymbol{\theta})}\right)  d\boldsymbol{x}. \label{K-L}%
\end{equation}
It is straightforward to check that $\mathcal{D}_{\mbox{\tiny KL}}%
[p_{\boldsymbol{\theta}},p_{\boldsymbol{\theta}^{\prime}}]\geq0$ and
$\mathcal{D}_{\mbox{\tiny KL}}[p_{\boldsymbol{\theta}},p_{\boldsymbol{\theta
}^{\prime}}]=0$ if and only if $p_{\boldsymbol{\theta}}=p_{\boldsymbol{\theta
}^{\prime}}$. In addition, by means of simple algebraic computations, it
follows that
\begin{equation}
\frac{\partial^{2}\mathcal{D}_{\mbox{\tiny KL}}}{\partial\theta^{i}%
\partial\theta^{j}}|_{\boldsymbol{\theta}=\boldsymbol{\theta}^{\prime}}%
=g_{ij}(\boldsymbol{\theta}),\ -\frac{\partial^{3}\mathcal{D}%
_{\mbox{\tiny KL}}}{\partial\theta^{i}\partial\theta^{j}\partial\theta^{\prime
k}}|_{\boldsymbol{\theta}=\boldsymbol{\theta}^{\prime}}=\overset{e}{\nabla
}_{ijk}\text{,}\text{ and}\ -\frac{\partial^{3}\mathcal{D}_{\mbox{\tiny KL}}%
}{\partial\theta^{\prime i}\partial\theta^{\prime^{j}}\partial\theta^{k}%
}|_{\boldsymbol{\theta}=\boldsymbol{\theta}^{\prime}}=\overset{e}{\nabla
^{\ast}}_{ijk}\text{.} \label{div-metric}%
\end{equation}
As previously mentioned, we refer to Refs. \cite{amari85,amari,amari16} for
additional mathematical details on IG.

\section{Information Geometry and Phase Transitions}

{In this section, we report on the use of IG methods to characterize classical
phase transitions of second-order that emerge in a variety of physical
scenarios. }

The information theoretic analysis of second-order phase transitions in
classical systems relies on the definition of a Riemannian metric on the
equilibrium thermodynamic state space of a system. In Ref. \cite{Rup95},
Ruppeiner conjectured that the scalar curvature $\mathcal{R}$ [Appendix
\ref{DiffGeom}] arising from such a metric is related to the correlation
volume of the system, i.e.,
\begin{equation}
\mathcal{R}\sim\xi^{d}\text{,} \label{achi}%
\end{equation}
where $\xi$ is the correlation length and $d$ is the system dimension. This
association follows from the idea that to a greater \textquotedblleft
distance\textquotedblright\ between two equilibrium thermodynamic states there
corresponds a smaller probability that these can be related by a fluctuation.
We remark that from a physics standpoint, the sign of $\mathcal{R}$ is linked
to the nature of the interparticle interactions in composite thermodynamical
systems \cite{Rup10}. Specifically, the sign of the scalar curvature regarded
as a measure of the strength of interaction can be zero, positive or negative
when such interactions are absent (that is, flat geometry for an ideal gas),
repulsive or attractive, respectively. The \textquotedblleft
distance\textquotedblright\ between two probability distributions in
parametric statistics can be evaluated, as mentioned in\ Section II, starting
from the Fisher-Rao information metric tensor of the system. In statistical
mechanics, the probability distributions $p(\boldsymbol{x};\boldsymbol{\theta
})$ of major interest are those exhibiting the so-called Gibbs form,
\begin{equation}
p(\boldsymbol{x};\boldsymbol{\theta})\overset{\text{def}}{=}\exp\left[
-\sum_{i=1}^{n}\theta^{i}\mathcal{H}_{i}(\boldsymbol{x})-\ln\mathcal{Z}%
(\boldsymbol{\theta})\right]  \text{,}%
\end{equation}
where $\boldsymbol{x}$ characterizes the state of the system (for instance,
spins), $\mathcal{H}_{i}(\boldsymbol{x})$ are the various Hamiltonian terms,
$\mathcal{Z}(\boldsymbol{\theta})$ is the normalizing partition function and,
finally, $\theta^{i}$ are the statistical parameters (for instance, the
inverse temperature $\beta$, the external magnetic field intensity $h$, and so
on). The $n$-dimensional manifold $\mathcal{M}$ of statistical parameters is
endowed with the Fisher-Rao information metric that measures the distance
between different probabilistic configurations. The components of the
Fisher-Rao information metric tensor can be expressed as,
\begin{equation}
g_{ij}\overset{\text{def}}{=}\partial_{i}\partial_{j}f\text{, } \label{GF}%
\end{equation}
where $f\overset{\text{def}}{=}\ln\mathcal{Z}$ is the reduced free energy per
site. For one-dimensional spin models, $\mathcal{M}$ is a two-dimensional
manifold with local coordinates $(\theta^{1},\theta^{2})=(\beta,h)$ and the
scalar curvature $\mathcal{R}$ is given by \cite{Rup95},
\begin{equation}
\mathcal{R}\overset{\text{def}}{=}-\frac{1}{2(\det g)^{2}}\det\left(
\begin{array}
[c]{ccc}%
\partial_{\beta}^{2}f & \partial_{\beta}\partial_{h}f & \partial_{h}^{2}f\\
&  & \\
\partial_{\beta}^{3}f & \partial_{\beta}^{2}\partial_{h}f & \partial_{\beta
}\partial_{h}^{2}f\\
&  & \\
\partial_{\beta}^{2}\partial_{h}f & \partial_{\beta}\partial_{h}^{2}f &
\partial_{h}^{3}f
\end{array}
\right)  \text{.} \label{curvature}%
\end{equation}
It is worth noticing that, unlike many statistical mechanical quantities, the
curvature $\mathcal{R}$ in Eq. (\ref{curvature}) depends on third order
derivatives of $f$. However, the consideration of $\mathcal{R}$ offers the
possibility of determining critical exponents in a non-standard way. Indeed,
the hypothesis that for a second-order phase transition the curvature depends
on the correlation volume combined with standard scaling assumptions and the
hyperscaling relation, $\nu d=2-\alpha$ \cite{janke02}, leads to
\begin{equation}
\mathcal{R}\sim|\beta-\beta_{c}|^{\alpha-2}\text{,} \label{scalingr}%
\end{equation}
where $\alpha$ is the exponent characterizing the scaling of the specific heat
while $\beta_{c}$ denotes the inverse of the critical temperature.

\subsection{The one-dimensional Ising model}

The scaling behavior in Eq. (\ref{scalingr}) near criticality of the scalar
curvature $\mathcal{R}$ of the two-dimensional parameter manifold was analyzed
in Ref. \cite{jany89} for the one-dimensional ($1D$) Ising model defined by
the Hamiltonian,
\begin{equation}
\mathcal{H}_{\text{{\tiny 1D -Ising}}}\overset{\text{def}}{=}-J\sum_{k=1}%
^{N}s_{k}s_{k+1}-h\sum_{k=1}^{N}s_{k}\text{,} \label{h1d}%
\end{equation}
where the spins $\left\{  s_{k}\right\}  _{k=1,...,N}$ are such that
$s_{k}=+1$ (spin up) or $s_{k}=-1$ (spin down), and $s_{N+k}=s_{k}$. Moreover,
$J$ is the coupling constant for the nearest-neighbor pairs in the lattice and
$h$ denotes the external magnetic field multiplied by the magnetic dipole
moment of one spin. The equilibrium probability distribution function $\rho$
of the $1D$ Ising model is given by,
\begin{equation}
\rho\overset{\text{def}}{=}\frac{\exp\left(  -\beta\mathcal{H}\right)
}{\mathcal{Z}\left(  \beta\text{, }\beta h\right)  }\text{,}%
\end{equation}
where $\mathcal{H}$ is defined in Eq. (\ref{h1d}), $\beta\overset{\text{def}%
}{=}T^{-1}$ (the Boltzmann constant is set equal to one in the manuscript),
and $\mathcal{Z}$ denotes the partition function,
\begin{equation}
\mathcal{Z}\left(  \beta\text{, }\beta h\right)  \overset{\text{def}}%
{=}e^{N\beta J}\left\{  \cosh\left(  \beta h\right)  +\left[  \cosh^{2}\left(
\beta h\right)  -2e^{-2\beta J}\sinh\left(  2\beta J\right)  \right]
^{\frac{1}{2}}\right\}  ^{N}\text{.} \label{zetat1}%
\end{equation}
The information geometry for this model was implemented\textbf{ }in the limit
of large $N$ in Ref. \cite{jany89}. In particular, its curvature
$\mathcal{R}_{\text{{\tiny 1D -Ising}}}$ is given by
\begin{equation}
\mathcal{R}_{\text{{\tiny 1D -Ising}}}=1+\left[  \sinh^{2}(\beta h)+e^{-4\beta
J}\right]  ^{-\frac{1}{2}}\cosh(\beta h)\text{.} \label{curvature1d}%
\end{equation}
It is interesting to note that $\mathcal{R}_{\text{{\tiny 1D -Ising}}}$ in Eq.
(\ref{curvature1d}) is a positive function of $\beta h$ and $\beta J$.

In particular, the scalar curvature does not depend on the orientation of the
external magnetic field $h$ since it is a symmetric function of the latter,
being $\mathcal{R}_{\text{1D -Ising}}\left(  h\right)  =\mathcal{R}_{\text{1D
-Ising}}\left(  -h\right)  $. The $1D$ Ising model can be thought of as having
a zero-temperature phase transition. Therefore, near the criticality regime
for the $1D$ Ising model, in the limit of $h=0$ and $\beta$ approaching
infinity, from Eq. (\ref{curvature1d}) one obtains (setting $J=1$)
\begin{equation}
\mathcal{R}_{1D\text{{\tiny -Ising}}}\sim e^{2\beta}\text{,} \label{ising1d}%
\end{equation}
corresponding to the expected $\alpha=1$ and $\nu=1$ with $\mathcal{R}%
_{1D\text{-Ising}}\sim\xi^{d}=\left\vert \xi\right\vert ^{\frac{2-\alpha}{\nu
}}$ and $\xi\overset{\text{def}}{=}-\ln\left[  \tanh\left(  \beta\right)
\right]  \sim e^{2\beta}$. \cite{dolan02}

\subsection{The one-dimensional $q$-state Potts model}

In analogy to the one-dimensional Ising model, it is possible to obtain the
expression of the scalar curvature $\mathcal{R}$ for the curved parameter
manifold that characterizes the one-dimensional $q$-state Potts model with
$q=2$ for the Ising model \cite{dolan02,Dol98}. The partition function of the
$1D$ $q$-state Potts model is defined as,
\begin{equation}
\mathcal{Z}_{N}\left(  y\text{, }z\right)  \overset{\text{def}}{=}%
\sum_{\left\{  s\right\}  }\exp\left\{  \beta\sum_{k=1}^{N}\left[
\delta\left(  s_{k}\text{, }s_{k+1}\right)  -\frac{1}{q}\right]  +h\sum
_{k=1}^{N}\left[  \delta\left(  s_{k}\text{, }1\right)  -\frac{1}{q}\right]
\right\}  \text{,}%
\end{equation}
with the spins $s_{k}\in\left\{  1\text{,..., }q\right\}  $. In this case, the
scalar curvature $R_{\text{{\tiny Potts}}}$ assumes the functional form,
\begin{equation}
\mathcal{R}_{\text{{\tiny Potts}}}\left(  q,e^{\beta},e^{h}\right)
\overset{\text{def}}{=}A\left(  q,e^{\beta},e^{h}\right)  +\frac{B\left(
q,e^{\beta},e^{h}\right)  }{\sqrt{\eta\left(  q,e^{\beta},e^{h}\right)  }%
}\text{,} \label{rpotts}%
\end{equation}
where $\eta\left(  q,e^{\beta},e^{h}\right)  $ is given by,
\begin{equation}
\eta\left(  q,e^{\beta},e^{h}\right)  \overset{\text{def}}{=}\left[  e^{\beta
}\left(  1-e^{h}\right)  +q-2\right]  ^{2}+\left(  q-1\right)  e^{h}\text{.}%
\end{equation}
The quantities $A$ and $B$ in Eq. (\ref{rpotts}) are smooth functions of
$e^{\beta}$ and $e^{h}$ and, importantly, do not diverge for finite values of
temperature and/or fields. Observe that Eq. (\ref{rpotts}) reduces to Eq.
\eqref{ising1d} for $q=2$. \cite{dolan02} Unlike the $1D$ Ising model, the
scalar curvature in Eq. (\ref{rpotts}) is no longer positive definite and the
$h\rightarrow-h$ symmetry is no longer present. Finally, despite the absence
of a true phase transition, in the the limit of $h=0$ and $\beta$ approaching
infinity, Eq. (\ref{rpotts}) yields the divergence of the scalar curvature
$\mathcal{R}_{\text{{\tiny Potts}}}$ at zero temperature,
\begin{equation}
\mathcal{R}_{\text{{\tiny Potts}}}\sim e^{\beta}\text{.} \label{potts}%
\end{equation}
For further details on the $1D$ $q$\emph{-}state Potts model, we refer to
Refs. \cite{dolan02,johnston03,janke04}.

\subsection{The Ising model on dynamical planar random graphs}

Within the framework of classical PTs, the examples of $1D$ Ising model and
$1D$ $q$-state Potts model exhibit unsatisfactory characteristics: the former
has no real phase transition, while the latter is mean field in nature. These
facts motivated the search for soluble models beyond mean-field theory and
with a genuine finite-temperature phase transition. One such case is
represented by the so-called Ising model on planar random graphs
\cite{janke02}. In Refs. \cite{kaza1,kaza2}, the first investigations of the
Ising model on an ensemble of $\Phi^{4}$ ($4$-regular) or $\Phi^{3}$
($3$-regular) planar random graphs were presented. These investigations were
motivated by string-theoretic considerations where the continuum limit of the
theory represents matter coupled to $2D$ quantum gravity \cite{kaza1,kaza2}.

In Ref. \cite{janke02}, the scaling behavior of the scalar curvature for the
Ising model on an ensemble of planar random graphs (that is, a graph in which
each possible edge is present or not with a certain probability) was
investigated from an information geometric perspective. The partition function
$\mathcal{Z}$ of this model is given by,
\begin{equation}
\mathcal{Z}\overset{\text{def}}{=}\sum_{n=1}^{\infty}\left[  \frac
{-4ge^{-2\beta}}{\left(  1-e^{-4\beta}\right)  ^{2}}\right]  ^{n}%
\mathcal{Z}_{n}\text{,}%
\end{equation}
where
\begin{equation}
\mathcal{Z}_{n}\overset{\text{def}}{=}\sum_{\left\{  G^{n}\right\}  }%
\sum_{\left\{  s\right\}  }\exp\left(  \beta\sum_{\left\langle
i,j\right\rangle }G_{ij}^{n}s_{i}s_{j}+h\sum_{i}s_{i}\right)  \ \text{,}%
\end{equation}
$g$ is the coupling constant, $G_{ij}^{n}$ is the connectivity matrix for an
$n$-vertex planar graph, and $G^{n}$ denotes a graph with $n$-vertices. {A
finding} of interest uncovered in this investigation is that the general
scaling behavior of the scalar curvature $\mathcal{R}_{\text{{\tiny Ising-PRG}%
}}$ is not recovered,
\begin{equation}
\mathcal{R}_{\text{{\tiny Ising-PRG}}}\sim\xi^{d}\sim\left\vert \beta
-\beta_{c}\right\vert ^{\alpha-2}\text{,} \label{yoyo1}%
\end{equation}
where, as pointed out earlier, $\xi$ is the correlation length, $d$ is the
dimensionality of the system, $\alpha$ is the critical exponent, and
$\beta_{c}$ denotes the critical value of $\beta$. Indeed, setting $h=0$,
omitting technical details that can be found in Ref. \cite{johnston03} , the
leading asymptotic term of the scalar curvature $\mathcal{R}%
_{\text{{\tiny Ising-PRG}}}$ becomes
\begin{equation}
\mathcal{R}_{\text{{\tiny Ising-PRG}}}\sim\frac{225}{176}\left\vert
\beta-\beta_{c}\right\vert ^{-2}\text{.} \label{yoyo2}%
\end{equation}
However, while $\alpha=1$ for the standard $1D$ Ising model, the standard
critical exponent for this model is $\alpha=-1$ \cite{kaza1,kaza2}. Therefore,
there is a discrepancy between Eq. (\ref{yoyo1}) with $\alpha=-1$ and Eq.
(\ref{yoyo2}). Such discrepancy was ascribed to the effect of the negativity
of the critical exponent on the components of the metric tensor and, as a
consequence, on the scalar curvature. For further details, we refer to Ref.
\cite{janke02}.

For the sake of completeness, we point out here that the spherical model of a
ferromagnet \cite{kac52} was also studied from an information geometric
perspective \cite{janke03}. In analogy to Ref. \cite{janke02}, the scalar
curvature of the this model exhibits a scaling behavior of the form
$\mathcal{R}_{\text{{\tiny Ising-PRG}}}\sim\left\vert \beta-\beta
_{c}\right\vert ^{-2}$, although the critical exponent $\alpha$ equals $-1$ in
this case \cite{kac52}. For further details, we refer to ref. \cite{janke03}.
Finally, we refer to Ref. \cite{brody03} for an intriguing investigation on
the physical interpretation of the scalar curvature of parameter manifolds
corresponding to finite Ising models.

\subsection{The Erd\"{o}s-R\'{e}nyi phase transition in uniform random graphs}

The Erd\"{o}s-R\'{e}nyi theorem \cite{ER60} proves the existence of a phase
transition undergone by random graphs. This is a classical example of an
analytically known major variation of the degree of complexity of a network.
It reveals in a rapid growth of the largest components merging suddenly in a
giant component, much larger than any of the remaining ones. In the next
section we will address this paradigmatic phase transition by making use of a
geometric entropy \cite{Franzosi15}.


\section{Geometric Entropy and Complex Networks}

In this section, we report on a Riemannian geometric entropy that is above all
an attempt to lift heterogeneous systems (complex networks) to homogeneous
systems (differential manifolds) \cite{Felice14}. More precisely, this
modelling scheme consists in considering random variables as sitting on the
vertices of a network and their correlations as weighted edges. Specifically,
consider a set of $n$ random variables $X_{1},\ldots,X_{n}$ distributed
according to a multivariate Gaussian probability distribution
$p(\boldsymbol{x};\boldsymbol{\theta})$ with zero mean value,
\begin{equation}
p(\boldsymbol{x};\boldsymbol{\theta})\overset{\text{def}}{=}\frac{\exp\left[
-\frac{1}{2}\boldsymbol{x}^{\top}\ \Sigma^{-1}\ \boldsymbol{x}\right]
}{\left[  (2\pi)^{n}\det\Sigma\right]  ^{1/2}} \label{Gausszeromean}%
\end{equation}
where $\boldsymbol{x}^{\top}=(x_{1},\ldots,x_{n})\in\mathbb{R}^{n}$, with
$\top$ denoting the transposition. In addition, $\boldsymbol{\theta}%
=(\theta^{1},\ldots,\theta^{m})$ are the real-valued parameters that
characterize the above probability distribution function in Eq.
(\ref{Gausszeromean}). Such parameters are the entries in the covariance
matrix $\Sigma$. In this case, the Gaussian model in Eq. (\ref{Gausszeromean})
happens to be an $m$-dimensional statistical model with $m=\frac{n(n+1)}{2}$.
The parameter space $\Theta$ is the space of the variances and covariances of
the above mentioned random variables; it is given by all $\boldsymbol{\theta
}\in\mathbb{R}^{n}$ such that $\Sigma(\boldsymbol{\theta})>0$ (that is,
$\Sigma$ is a positive definite matrix). Now, the space $\Theta$ is endowed
with the Fisher-Rao information metric tensor $g=g_{ij}d\theta^{i}\otimes
d\theta^{j}$ whose components {are defined in \eqref{F-R}.}

In this way, a configuration space is associated to each network. Such a space
consists of a subset of the linear vector space $\mathbb{R}^{m}$ given by the
$m$-parameters that characterize the joint probability distribution of the
random variables. In turn, this configuration space is endowed with the
Fisher-Rao information metric. Then, in analogy to classical statistical
mechanics, a Riemannian geometric entropy is defined as the logarithm of the
volume of this {manifold}
\begin{equation}
\mathcal{S}\overset{\text{def}}{=}\ln\mathcal{V}_{ \boldsymbol{\theta}%
}\text{,} \label{entropygeneral}%
\end{equation}
where $\mathcal{V}_{\boldsymbol{\theta}}$ is the Riemannian volume of the
parameter space $\Theta$.

{The fact that this quantity was proposed in the context of networks inspired
its analysis in the Erd\"{o}s-R\'{e}nyi phase transition in random} graph{
\cite{ER60}.} A random graphs model $\mathbb{G}(n,k)$ is devised by choosing
with uniform probability a graph from the set of all graphs having $n$
vertices and $k$ edges \cite{Lucz00}. We can think of a type of dynamics by
adding the edges one at a time. When $k$ has the same order of magnitude as
$n$, the evolution from $k=0$ to $k=\binom{n}{k}$ yields, according to the
Erd\"{o}s-R\'{e}nyi theorem \cite{ER60}, a phase transition, revealing itself
in a rapid growth with $k$ of the size of the largest component. Specifically,
the structure of graphs when the expected degree of each of its vertices is
close to $1$, i.e. $k\sim n/2$, shows a jump: the order of magnitude of the
size of the largest component of graphs rapidly grows from $\ln n$ to $n$, if
$k$ has the same order of magnitude of $n$.

{In Ref. \cite{Franzosi15}, the Riemannian geometric entropy $\mathcal{S}(k)$,
defined in Eq. (\ref{entropygeneral}), has been numerically evaluated as a
function of $k$ for a fixed $n$, in order to investigate its sensitivity to
the appearance of the giant component during the evolution of the random graph
model $\mathbb{G}(n,k)$.} Two main results were uncovered. First, in agreement
with typical features that arise in numerical investigations of second order
phase transitions, what asymptotically would be a sharp bifurcation is rounded
at finite $n$, in analogy to finite-size effects on the order parameter.
Second, considering
\begin{equation}
{\widetilde{\mathcal{S}}(k)}\overset{\text{def}}{{=}}\frac{1}{n}\left\langle
\mathcal{S}(k)-\mathcal{S}(0)\right\rangle \text{,} \label{nicco}%
\end{equation}
where $\left\langle \cdot\right\rangle $ denotes a Monte Carlo estimation of
the average, it was observed in Ref. \cite{Franzosi15} that the larger $n$ is,
the more pronounced the \textquotedblleft knee\textquotedblright\ of
${\widetilde{\mathcal{S}}(k/n)}$ becomes. Interestingly, this finding is in
agreement with an $n$-asymptotic bifurcation at $k/n=0.5$ where the phase
transition takes place.

This unprecedented result suggests to propose the Riemannian geometric entropy
as a measure of networks complexity. {To this end let us recall that the
\textquotedblleft state manifold\textquotedblright\ $\mathcal{M}%
\overset{\text{def}}{=}(\Theta,g)$ associated with a given network of $n$
nodes has, in principle, dimension $m=\frac{n(n+1)}{2}$. Then, in order to
simplify the computation of components of }the Fisher-Rao {information
metric,} we define a new Riemannian metric which accounts also for the network
structure given by the adjacency matrix \cite{Franzosi15}. We first consider a
trivial network with null adjacency matrix and interpret $n$ independent
Gaussian variables $X_{1},\ldots,X_{n}$ as sitting on its vertices. It is
clear that the joint probability distribution is now given by
\eqref{Gausszeromean} with covariance matrix as $\Sigma_{0}%
=\mbox{\rm diag}[\theta^{n},\ldots,\theta^{n}]$, where $\theta^{i}%
=\mathbb{E}[X_{i}^{2}]$ for all $i:1,\ldots,n$. In this case, the Riemannian
manifold $\mathcal{M}_{0}\overset{\text{def}}{=}(\Theta_{0},g_{0})$ associated
with the bare network is given by \cite{Felice14}
\begin{align}
&  \Theta_{0}\overset{\text{def}}{=}\{\boldsymbol{\theta}=(\theta^{1}%
,\ldots,\theta^{n})\ |\ \theta^{i}>0\},\nonumber\\
&  g_{0}\overset{\text{def}}{=}\frac{1}{2}\sum_{i=1}^{n}\ \left(  \frac
{1}{\theta^{i}}\right)  ^{2}d\theta^{i}\otimes d\theta^{i}\ \text{.}
\label{bare}%
\end{align}
From \eqref{bare}, it is clear that a functional relation holds between the
components of the metric $g_{0}$ and the entries of the matrix $\Sigma_{0}$.
This is given by $g_{ii}=\frac{1}{2}\left(  \sigma_{ii}^{-1}\right)  ^{2}$
where $\sigma_{ii}^{-1}$ is the $ii$-th entry of the inverse matrix of
$\Sigma_{0}$, i.e. $\sigma_{ii}^{-1}=\frac{1}{\theta^{i}}$. Inspired by this
functional relation, we associate a Riemannian manifold to any network
$\mathcal{X}$ with non vanishing adjacency matrix $A$. To this aim, consider
the map $\psi_{\boldsymbol{\theta}}(A):\mbox{\rm A}(n,\mathbb{R}%
)\rightarrow\mbox{\rm GL}(n,\mathbb{R})$ defined by the relation
\begin{equation}
\psi_{\boldsymbol{\theta}}(A)\overset{\text{def}}{=}\Sigma_{0}%
(\boldsymbol{\theta})+A, \label{psi}%
\end{equation}
where $\mbox{\rm A}(n,\mathbb{R})$ denotes the set of symmetric $n\times n$
matrices over $\mathbb{R}$ with vanishing diagonal elements that can represent
any simple undirected graph. Here, $\mbox{\rm GL}(n,\mathbb{R})$ is the
general linear group{ of matrices}. Then, we deform the manifold $\mathcal{M}$
via $\psi_{\boldsymbol{\theta}}(A)$. Hence the manifold associated to a
network $\mathcal{X}$ with adjacency matrix $A$ is $\widetilde{\mathcal{M}%
}\overset{\text{def}}{=}(\widetilde{\Theta},\widetilde{g})$ where,
\begin{equation}
\widetilde{\Theta}\overset{\text{def}}{=}\{\boldsymbol{\theta}\in
\mathbb{R}\ |\ \psi_{\boldsymbol{\theta}}%
(A)\ \mbox{is positive definite}\}\text{,} \label{Riemvary}%
\end{equation}
and $\widetilde{g}\overset{\text{def}}{=}\widetilde{g}_{ij}d\theta^{i}\otimes
d\theta^{j}$. The components of $\widetilde{g}$ are given by,
\begin{equation}
\widetilde{g}_{ij}\overset{\text{def}}{=}\frac{1}{2}\left(  \psi
_{\boldsymbol{\theta}}^{-1}(A)_{ij}\right)  ^{2}\text{,} \label{F-Rvary}%
\end{equation}
where $\psi_{\boldsymbol{\theta}}^{-1}(A)_{ij}$ is the $ij$ entry of the
inverse matrix of $\psi_{\boldsymbol{\theta}}(A)$.

In this way, a differentiable system (Riemannian manifold) is associated to a
discrete system (network) through the description of the network by a set of
probability distribution functions. Now, having a phase space for describing
the network structure, we supply the microcanonical definition of entropy in
statistical mechanics in order to introduce a geometric entropy of a network
$\mathcal{X}$ with adjacency matrix $A$. It is given by
\begin{equation}
\mathcal{S}\overset{\text{def}}{=}\ln\mathcal{V}(A)\text{,} \label{entropy}%
\end{equation}
where $\mathcal{V}(A)$ is the volume of $\widetilde{\mathcal{M}}$ computed
from the volume element $\nu_{\widetilde{g}}$ defined as,
\begin{equation}
\nu_{\widetilde{g}}\overset{\text{def}}{=}\sqrt{\det\widetilde{g}}%
\ d\theta^{1}\wedge\ldots\wedge d\theta^{n}\text{,} \label{volelement}%
\end{equation}
where \textquotedblleft$\wedge$\textquotedblright\ in Eq. (\ref{volelement})
denotes the wedge product of $n$ $1$\textbf{-}forms. The volume $\mathcal{V}%
(A)$ in Eq. (\ref{entropy}) however, is ill-defined. This happens because
$\widetilde{\Theta}$ is not compact since the variables $\theta^{i}$ are
unbound from above. In addition, $\det\left[  \widetilde{g}(\boldsymbol{\theta
})\right]  $ can be zero for some $\theta^{i}$s and this causes the volume
element to diverge as it is clear from Eq. \eqref{F-Rvary}. Thus, as is
customary in the literature \cite{Leibb75}, we regularize $\mathcal{V}(A)$ as
follows
\begin{equation}
\mathcal{V}(A)=\int_{\widetilde{\Theta}}\mathcal{R}(\psi_{\boldsymbol{\theta}%
}(A))\,\nu_{\widetilde{g}}, \label{regular}%
\end{equation}
where $\mathcal{R}(\psi_{\boldsymbol{\theta}}(A))$ is any suitable
regularizing function. This procedure is essentially the compactification of
the parameter space where we exclude those sets of $\theta^{i}$ values which
make $\det\left[  \widetilde{g}(\boldsymbol{\theta})\right]  $ divergent.

Theoretically, a regularizing function $\mathcal{R}(\psi_{\boldsymbol{\theta}%
}(A))$ could be devised by taking into account the probabilistic nature of the
model description. Then, in order to make a natural choice for $\mathcal{R}%
(\psi_{\boldsymbol{\theta}}(A))$, consider the Shannon entropy
$H_{\text{Shannon}}$ of a multivariate normal distribution as the one in
Eq.\eqref{Gausszeromean},
\begin{equation}
H_{\text{Shannon}}=\ln\sqrt{(2\pi e)^{n}\det\Sigma(\boldsymbol{\theta}%
)}\text{.}%
\end{equation}
Interpreting the Shannon entropy as a measure of the lack of information about
a system, to a smaller variance of a Gaussian pdf there corresponds a smaller
entropy. In particular, in the limiting case in which\textbf{ }$\det
\Sigma(\theta)$\textbf{ }approaches zero for some\textbf{ }$\theta^{i}%
$\textbf{, }we have the minimum entropy and, consequently, the maximum
information about a system. In such a case, it is clearly meaningless to
consider a complexity measure. Similarly, if $\theta^{i}$ approaches infinity,
the Shannon entropy diverges and there is a lack of information. For these
reasons, it appears quite natural {to choose} the regularizing function
$\mathcal{R}(\psi_{\boldsymbol{\theta}}(A))$ as follows,%
\begin{align}
\mathcal{R}(\psi_{\boldsymbol{\theta}}(A))\overset{\text{def}}{=}  &  \left[
H(N-\mathrm{tr}\psi_{\boldsymbol{\theta}}(A))+H(\mathrm{tr}\psi
_{\boldsymbol{\theta}}(A)-N)e^{-\mathrm{tr}\psi_{\boldsymbol{\theta}}%
(A)}\right]  {\times}\nonumber\\
& \label{regfun1}\\
&  \times\left[  H(\varepsilon-\det\psi_{\boldsymbol{\theta}}(A))\ln\left[
1+\left(  \det\psi_{\boldsymbol{\theta}}(A)\right)  ^{n}\right]  +H(\det
\psi_{\boldsymbol{\theta}}(A)-\varepsilon)\right]  ,\nonumber
\end{align}
where $H(\cdot)$ is the Heaviside step function, $N$ denotes a positive {real}
number with $\mathcal{R}(\psi_{\boldsymbol{\theta}}(A))\sim{\ O}%
(\mathrm{tr}\psi_{\boldsymbol{\theta}}(A))$ for all $\boldsymbol{\theta}$ such
that $\mathrm{tr}\psi_{\boldsymbol{\theta}}(A)\geq N$, and $\varepsilon$ is a
positive \textit{real} number such that $\mathcal{R}(\psi_{\boldsymbol{\theta
}}(A))\sim{\ o}(\det\psi_{\boldsymbol{\theta}}(A))$ for all
$\boldsymbol{\theta}$ such that $\det\psi_{\boldsymbol{\theta}}(A)\leq
\varepsilon$. It happens that the volume $\mathcal{V}(A)$ in Eq.
(\ref{regular}) is well-defined \cite{Felice17}.

The geometric entropy \eqref{regular} is used to single out emergent phenomena
occurring in power-law random graphs, in addition to uniform random graphs.
The former are described by two parameters, $\alpha$ and $\gamma$, which
define the size and the density of a network; hence, given the number of nodes
$n$ with degree $d$, this model, denoted $\mathbb{G}_{\gamma,d}$
\cite{Latora06}, assigns a uniform probability to all graphs with
$n=e^{\alpha}d^{-\gamma}$ where $e$ represents the Euler number. Now,
according to the $n$-asymptotic bifurcation at $\gamma=\gamma_{c}=3.47875$
predicted by the Molloy-Reed criterion for the emergence of a giant component
in $\mathbb{G}_{\gamma,d}$ \cite{Molloy}, the geometric entropy
\eqref{regular} displays the typical phenomenon uncovered in numerical
investigations of second-order phase transitions when $\gamma$ is in the range
$1.5<\gamma<5.5$ \cite{franzosi16}.

In the transition from random graphs to real networks, the geometric entropy
so far put forward has been considered against some small exponential random
graphs which are evaluated by \eqref{AyComplMeas} in order to describe
\textquotedblleft typical\textquotedblright\ graphs, i.e., the graphs that are
most probable in the ensemble defined by this model and that correspond to the
lowest \textquotedblleft energy\textquotedblright\ characterizing the model
\cite{olbrich08}. In Ref. \cite{olbrich08}, the convex hull of all possible
expectation values of the probabilities of the triangles and $3$-chains is
derived. Furthermore, graphs that correspond to the minimal energy are found
to lie on the lower boundary of the mentioned convex hull. The geometric
entropy \eqref{regular} suggests that going up along the lower boundary of
this convex hull, the degree of complexity increases \cite{franzosi16}.
However, for the family of graphs which are intended of maximal energy in Ref.
\cite{olbrich08}, it happens that network complexity is nontrivially
influenced by network topology (homology) as was first pointed out in Ref.
\cite{Felice14}.

An additional step towards analyzing real networks is to consider $d$-regular
graphs \cite{Lucz00}. They are networks where each node has the same degree
$d$. A way to generate these special networks is the \emph{configuration
model} \cite{Bender78} that is specified in terms of a sequence of degrees
$d=(d_{1},\ldots,d_{n})$. The average vertex degree $\langle d_{i}\rangle$ is
the ratio between the total number of links in a given network and the number
of nodes. It represents the first level of characterization of the topological
complexity \cite{Bender78}. According to the geometric entropy, the larger
$d$, the more complex the network is. Similarly, when we consider networks
with one or more hubs, i.e. nodes with degree larger than the others, the
complexity of a network increases with the number of hubs in it
\cite{franzosi16}. Now, $d$-regular graphs show homogeneity in the interaction
structures entailing topological equivalence of almost all the nodes. On the
contrary, most real networks have power law degree distribution
$p(d)=\mathcal{A}\ d^{-\gamma}$, where $\mathcal{A}$ is a positive
\textit{real} constant and the exponent $\gamma$ varies in the range
$2\leq\gamma\leq3$ \cite{Latora06}, and are assessed with high heterogeneity,
that is a measure of how far from $p(d_{0})=1$, $p(d)=1(d_{0}\neq d)$ a
network is \cite{Wu2008}. The degree of heterogeneity $h$ of a network has
been quantified in \cite{Estrada10} by means of the following definition,
\begin{equation}
h\overset{\text{def}}{=}\sum_{i,j\in E}\left(  \frac{1}{\sqrt{d_{i}}}-\frac
{1}{\sqrt{d_{j}}}\right)  \text{,} \label{heterogeneity}%
\end{equation}
where $E$ is the set of edges of the network. This heterogeneity measure
vanishes for regular graphs, while, as the difference in the degrees of
adjacent nodes increases, it also increases. Interestingly, the degree of
heterogeneity of some networks as measured by \eqref{regular} is the same as
the ordering produced by the $h$ measure of heterogeneity \cite{franzosi16}.

Finally, the geometric entropy has been applied to some real networks where
its predicted features have been compared with those of three other measures
of complexity: one describing modularity structure, one homogeneity, and,
finally, one characterizing the heterogeneity of real networks \cite{kim08}.
The findings uncovered in these investigations validate the meaningfulness and
the effectiveness of the geometric entropy as a good measure of network
complexity \cite{franzosi16}.


\section{Stochastic interactions}

{Other complexity measures can be placed in a geometric framework.} When
considering the statistical (or, structural) complexity \cite{crutchfield89}
of describing a system made up by an ensemble of interacting units, a
reasonable feature of complex systems is the emergence of new structural
properties at different (more detailed) levels of description. Starting from
this observation, Ay and coworkers exploited the main results presented in
Ref. \cite{amari01,ay02} (hierarchy of probability distributions) and
\cite{darroch80} (hierarchical graphical models) in order to place well-known
complexity measures \cite{tononi94,grassberger86,olbrich08} in a unifying
information geometric setting. Within this IG framework, the idea of
hierarchical structure of probability distributions is used to characterize
the interactions among the units of a complex system at different scales of
description by means of an explicit orthogonal decomposition of the stochastic
dependency with respect to an interaction hierarchy (or, order) \cite{ay06}.
In Ref. \cite{Aypre}, Ay and coworkers introduced a notion of complexity based
on the interaction among parts of a system that rigorously incorporates the
idea to quantify the amount of pairwise or triplewise interacting components
and beyond, by using the exponential families of $k$-interactions and the
notion of distance to such an exponential family. This leads to quantify
complexity in terms of the degree of $k$-wise interaction that cannot be
explained by a $(k-1)$-wise interaction. For further details on this approach
by Ay and coworkers, we refer to Refs. \cite{ay11,ay15}. In particular, for an
investigation on the effects of both spatial and temporal interdependencies in
complex systems of stochastically interacting units, we refer to Ref.
\cite{ay15}.

In what follows, we shall present the essential tools needed to understand the
geometric complexity measure introduced by Ay and coworkers.

\subsection{Exponential families of $k$-interactions}

The set of states of a system is given in terms of compositional structure,
i.e. for each site index $v$ in a set $V\overset{\text{def}}{=}\{1,\ldots,N\}$
we have a finite configuration space $\mathcal{X}_{v}$. The set of all
possible configuration is $\mathcal{X}_{V}\overset{\text{def}}{=}\times_{v\in
V}\mathcal{X}_{v}$ and likewise $\mathcal{X}_{A}\overset{\text{def}}{=}%
\times_{v\in A}\mathcal{X}_{v}$ for each subset $A\subset V$. Given the set of
real-valued functions on $\mathcal{X}_{v}$, $\mathbb{R}^{\mathcal{X}_{v}%
}\overset{\text{def}}{=}\{f\ |f:\mathcal{X}_{v}\rightarrow\mathbb{R}\}$,
elements of the subset $\mathcal{F}_{\mathcal{X}_{V}}$,
\begin{equation}
\mathcal{F}_{\mathcal{X}_{V}}\overset{\text{def}}{=}\left\{  p\in
\mathbb{R}^{\mathcal{X}_{V}}:p\left(  \boldsymbol{x}\right)  \geq0\text{,
}\underset{\boldsymbol{x}\in\mathcal{X}_{V}}{\sum}p\left(  \boldsymbol{x}%
\right)  =1\right\}  \text{,} \label{fv}%
\end{equation}
are probability measures on $\mathcal{X}_{v}$. Indeed, due to the
compositional structure, these probability measures are only joint
probabilities of a set of random variables $\{X_{v}\ |v\in V\}$, where $X_{v}$
takes values in $\mathcal{X}_{v}$. Observe that from a geometrical viewpoint,
$\mathcal{F}_{\mathcal{X}_{V}}$ in Eq. (\ref{fv}) can be regarded as a
$\left(  \left\vert \mathcal{X}_{V}\right\vert -1\right)  $-dimensional
probability simplex. Now, the support of a probability distribution $p$ is
given by,
\begin{equation}
\text{\emph{supp}}\left(  p\right)  \overset{\text{def}}{=}\left\{
\boldsymbol{x}\in\mathcal{X}_{V}:p\left(  \boldsymbol{x}\right)  >0\right\}
\text{.}%
\end{equation}
and probability distributions with full support are denoted as $\mathcal{P}%
\left(  \mathcal{X}_{V}\right)  $,
\begin{equation}
\mathcal{P}\left(  \mathcal{X}_{V}\right)  \overset{\text{def}}{=}\left\{
p\in\mathcal{F}_{\mathcal{X}_{V}}:p\left(  \boldsymbol{x}\right)  >0\text{,
}\forall\boldsymbol{x}\in\mathcal{X}_{V}\right\}  \text{.}%
\end{equation}
In order to define useful exponential families for quantifying the amount of
$k$-interactions, we consider the linear space of functions depending on only
$k$ of their arguments,
\begin{equation}
\mathcal{I}_{k}\overset{\text{def}}{=}\underset{A\subseteq V\text{,
}\left\vert A\right\vert =k}{\emph{span}}\left(  \mathcal{I}_{A}\right)
\text{,}%
\end{equation}
where $\mathcal{I}_{A}$ is a subspace of $\mathbb{R}^{\mathcal{X}_{V}}$ formed
by all the functions that do not depend on the configurations outside $A$,
\begin{equation}
\mathcal{I}_{A}\overset{\text{def}}{=}\left\{  f\in\mathbb{R}^{\mathcal{X}%
_{V}}:f\left(  \boldsymbol{x}_{A}\text{, }\boldsymbol{x}_{V\backslash
A}\right)  =f\left(  \boldsymbol{x}_{A}\text{, }\widetilde{\boldsymbol{x}%
}_{V\backslash A}\right)  \right\}  \text{,}%
\end{equation}
for all $\boldsymbol{x}_{A}\in\mathcal{X}_{A}$, $\boldsymbol{x}_{V\backslash
A}$ and $\widetilde{\boldsymbol{x}}\in\mathcal{X}_{V\backslash A}$. After
taking the exponential map of sums of such functions, we find the probability
distributions with only $k$ interactions $\mathcal{E}_{k}\overset{\text{def}%
}{=}\exp\left(  \mathcal{I}_{k}\right)  $, where
\begin{equation}
\exp:\mathbb{R}^{\mathcal{X}_{V}}\ni f\left(  x\right)  \mapsto\frac
{\exp\left[  f\left(  x\right)  \right]  }{\underset{\boldsymbol{x}%
\in\mathcal{X}_{V}}{\sum}\exp\left[  f\left(  x\right)  \right]  }%
\in\mathcal{P}\left(  \mathcal{X}_{V}\right)  \text{.}%
\end{equation}
Proceeding along these lines of reasoning, a hierarchy of exponential families
can be defined in terms of the following chain of inclusions,
\begin{equation}
\mathcal{E}_{1}\subseteq\mathcal{E}_{2}\ldots\subseteq\mathcal{E}_{k}\text{.}
\label{hierexp}%
\end{equation}
This hierarchy was widely studied in Refs. \cite{amari01,AyKnauf} where
various applications in the theory of neural networks were considered. The
connection between this formalism and statistical physics relies upon the
following correspondence: $f\in\mathcal{I}_{k}$ corresponds to an interaction
energy which has only $k$ interactions, but no higher interactions. This fact
gives rise to a probability distribution $p(\boldsymbol{x})$,
\begin{equation}
p(\boldsymbol{x})\overset{\text{def}}{=}\frac{\exp\left[  f\left(
\boldsymbol{x}\right)  \right]  }{\mathcal{Z}}\in\mathcal{E}_{k}\text{,}
\label{Stat}%
\end{equation}
where $\mathcal{Z}$ denotes the partition function of the system and it is a
normalization factor. Let us note that, in general, the image of the
exponential map is given by distributions with full support. Therefore,
probability zero, which corresponds to infinite energy, is achievable only by
limits of sequences of probability measures in the exponential family. This is
not a serious drawback for considering the Kullback-Leibler divergence for
introducing a complexity measure based on the notion of distance to an
exponential family $\mathcal{E}$.

\subsection{Relative entropy and interaction structures}

According to Eq. \eqref{K-L} the Kullback-Leibler divergence (or, relative
entropy) $\mathcal{D}\left(  p||q\right)  $ describes the distance between two
probability distributions $p$ and $q$ in $\mathcal{F}_{\mathcal{X}_{V}}$ even
though it is not a metric; it is not a metric since it is not symmetric and
does not satisfy the triangle inequality. As a consequence, one can consider
the distance to an exponential family\textbf{ }$\mathcal{E}$\textbf{ }defined
in terms of the Kullback-Leibler divergence as,
\begin{equation}
\mathcal{D}\left(  p||\mathcal{E}\right)  \overset{\text{def}}{=}\inf
_{q\in\mathcal{E}}\mathcal{D}\left(  p||q\right)  \text{,} \label{expdist}%
\end{equation}
which satisfies the following relation,
\begin{equation}
\mathcal{D}(p\Vert\mathcal{E})-\mathcal{D}(p\Vert\mathcal{F})\geq0
\end{equation}
whenever $\mathcal{E}\subseteq\mathcal{F}$. At this point, we are ready to
describe the complexity measure proposed by Ay and coworkers. This measure is
a vector-valued quantity $I\left(  p\right)  \overset{\text{def}}{=}\left(
I^{\left(  1\right)  }\left(  p\right)  \text{,..., }I^{\left(  N\right)
}\left(  p\right)  \right)  $ with,
\begin{equation}
I^{\left(  k\right)  }\left(  p\right)  \overset{\text{def}}{=}\mathcal{D}%
\left(  p||\mathcal{E}_{k-1}\right)  -\mathcal{D}\left(  p||\mathcal{E}%
_{k}\right)  \text{,} \label{AyComplMeas}%
\end{equation}
for any $1\leq k\leq N$. The component $I^{\left(  k\right)  }\left(
p\right)  $ is a quantitative measure of the dependencies among $k$ nodes that
fail to be captured by interaction among smaller subsets of nodes. An explicit
illustrative example of this nice theoretical approach to complexity appears
in Ref. \cite{Aypre} where, by considering a one-dimensional binary cellular
automata and the symbolic dynamics of a coupled tent map lattice
\cite{pethel06,cafarochaos}, a numerical analysis of such information
geometric complexity measures was presented. In particular, for elementary
binary cellular automata \cite{Wolf} with rules uniformly distributed, no
interaction can be detected. On the contrary, the complexity measure in the
Eq. \eqref{AyComplMeas} is able to detect long-range correlations for rules
that do not have the uniform distribution as their invariant measure. As far
as the coupled tent maps are concerned, the fully connected graph and the
circle graph are considered. For $k>6$, the $I^{(k)}$ quantities are very
small and depend on the random initial condition, so they are neglected. For
the fully connected graphs, the $I^{(k)}$ suggest that complex dynamics takes
place on the edge of synchronized regimes.\textbf{ }Furthermore, interesting
results are uncovered when the simulation of the symbolic dynamics of a
coupled tent map lattice is performed by means of a circle graph specified by
a next neighbor coupling. Specifically, when two nodes of the lattice are
constant and four of them are periodic, all\textbf{ }$I^{(k)}$\textbf{
}with\textbf{ }$k\neq2$\textbf{ }are zero.\textbf{ }This is a signature of a
high degree of regularity. However, when the average activity does not follow
one of the individual nodes and a complex dynamical structure driven by the
higher-order correlations among the nodes exists, very high values of
$I^{(k)}$ with $k\neq2$ are observed.

Relying on the correspondence with statistical physics described by Eq.
\eqref{Stat}, the complexity measure \eqref{AyComplMeas} is also able to
quantify structure in networks \cite{Olbrich2010}. Indeed, whenever the
function $f(\boldsymbol{x})$ is an Hamiltonian of the observables of the
system, the exponential family happens to be the collection of maximum entropy
distributions for fixed expectation values of these observables. The counts of
subgraphs with no more than $k$ links is a sufficient statistic for
hierarchically structured exponential families. Hence, within this theoretical
framework, the study of the dependencies among several observables such as the
degree distribution, cluster, and assortativity coefficients, leads to
quantifiable structures in networks.


\section{Entropic motion}

{We now move to temporal complexity arising in entropic inference processes.}
In recent years, entropic inference methods \cite{caticha12} have been
combined with information geometric techniques \cite{amari} in order to
quantify the complexity of statistical models used to provide probabilistic
descriptions of physical systems in the presence of limited information.
Within this context, the complexity of statistical models is regarded as the
difficulty of inferring macroscopic predictions in the absence of detailed
knowledge about the microscopic nature of the system being investigated. This
fascinating line of research, originally called \emph{Information Geometric
Approach to Chaos }(IGAC), was presented by Cafaro in his Ph.\ D. doctoral
dissertation in Ref. \cite{cafarophd}. The essence of the IGAC\ theoretical
construct can be described as follows: given an arbitrary complex system, once
the microscopic degrees of freedom of the system are identified and the
relevant information constraints about them are selected, entropic methods are
used to provide a preliminary static statistical model of the complex system.
The statistical model is defined by probability distributions parametrized in
terms of statistical macrovariables which, in turn, depend on the particular
functional form of the information constraints that have been assumed to be
relevant for carrying out the statistical inferences. Then, assuming that the
complex system evolves, the evolution of the corresponding statistical model
from an initial to a final configuration is described by means of the
so-called Entropic Dynamics (ED \cite{catichaED}). The ED is a form of
information-constrained dynamics built on curved statistical manifolds where
elements of the manifold are probability distributions that are in a
one-to-one relation with a suitable set of statistical macrovariables that
define the so-called parameter space and provide a convenient parametrization
of points on the statistical manifold. Within the ED framework, the evolution
of probability distributions is governed by an entropic inference principle:
starting from the initial configuration, motion towards the final
configuration occurs by maximizing the logarithmic relative entropy (Maximum
relative Entropy method- MrE method, \cite{caticha12}) between any two
consecutive intermediate configurations. It is worthwhile mentioning that
ED\ only provides the \emph{expected} and not the \emph{actual} trajectories
of the system \cite{cafaropre}. Inferences in ED rely on the nature of the
chosen information constraints employed in the MrE algorithm. The validation
of this type of modeling scheme can be checked only \emph{a posteriori}. If
discrepancies between the experimental observations and the inferred
predictions occur, a new set of information constraints needs to be selected.
The evolution of probability distributions is described in terms of the
outcome of this maximization procedure, namely a geodesic evolution of the
statistical macrovariables \cite{caticha12}. The distance between two
different probability distributions can be viewed as a measure of the
distinguishability between the two distributions and is quantified, as
mentioned in Section II, in terms of the so-called Fisher-Rao information
metric \cite{amari}. Given the information metric, {the evolution} arising{
from the maximization procedure results in geodesics with respect to the
Levi-Civita (or Riemannian) connection $\overline{\nabla}$ of the Fisher-Rao
metric. Now, the Riemannian connection $\overline{\nabla}$ gives {rise} to a
self dual structure of the statistical model meaning that $\nabla^{\ast
}=\nabla$. Within the IGAC approach }to complexity, the{ complete geometric
structure supplying the entropic dynamics is given in terms of the Fisher-Rao
metric and the Riemannian connection $\overline{\nabla}$ }due to minimization
requirements on the Lagrangian metric functional \cite{cafarophd}. {For this
reason, within the }present{ section, we refer to statistical models endowed
with a Riemannian geometric structure. In this way} one can apply standard
methods of Riemannian differential geometry to study the geometric structure
of the manifold underlying the entropic motion that characterizes the
evolution of probability distributions. Generally speaking, conventional
Riemannian geometric quantities such as Christoffel connection coefficients,
Ricci tensor, Riemannian curvature tensor, sectional curvatures, Ricci scalar
curvature , Weyl anisotropy tensor, Killing fields, and Jacobi fields can be
computed in the conventional manner. More specifically, the chaoticity (or, in
other words, the temporal complexity) of such statistical models can be
investigated using suitably chosen measures such as the signs of the scalar
and sectional curvatures of the statistical manifold underlying the entropic
motion, the asymptotic temporal behavior of Jacobi fields, the IG analogue of
Lyapunov exponents, the existence of Killing vectors, and the existence of a
non-vanishing Weyl anisotropy tensor. In addition to the above mentioned
ordinary measures of complexity imported into the information geometric
framework from the ordinary Riemannian geometry platform, complexity within
the IGAC\ approach can also be characterized in terms of the so-called
information geometric entropy (IGE), originally presented in \cite{cafarophd}.

\subsection{Information geometric entropy}

In what follows, we describe the concept of the IGE within the IGAC framework.
Let us assume that the points $\left\{  p\left(  \boldsymbol{x}%
;\boldsymbol{\theta}\right)  \right\}  $ of an $n$-dimensional curved
statistical manifold $\mathcal{M}_{s}$ are parametrized by means of $n$
\emph{real} valued variables $\left(  \theta^{1}\text{,..., }\theta
^{n}\right)  $,
\begin{equation}
\mathcal{M}_{s}\overset{\text{def}}{=}\left\{  p\left(  \boldsymbol{x}%
;\boldsymbol{\theta}\right)  :\boldsymbol{\theta}=\left(  \theta
^{1}\text{,..., }\theta^{n}\right)  \in\mathcal{D}_{\boldsymbol{\theta}%
}^{\left(  \text{tot}\right)  }\right\}  \text{.}%
\end{equation}
The microvariables $\boldsymbol{x}$ belong to the microspace $\mathcal{X}$
while the macrovariables $\boldsymbol{\theta}$ belong to the parameter space
$\mathcal{D}_{\boldsymbol{\theta}}^{\left(  \text{tot}\right)  }$ defined as,
\begin{equation}
\mathcal{D}_{\boldsymbol{\theta}}^{\left(  \text{tot}\right)  }\overset
{\text{def}}{=}{\displaystyle\bigotimes\limits_{k=1}^{n}}\mathcal{I}%
_{\theta^{k}}=\left(  \mathcal{I}_{\theta^{1}}\otimes\mathcal{I}_{\theta^{2}%
}\text{...}\otimes\mathcal{I}_{\theta^{n}}\right)  \subseteq\mathbb{R}%
^{n}\text{.} \label{dtot}%
\end{equation}
The quantity $\mathcal{I}_{\theta^{k}}$ with $1\leq k\leq n$ in Eq.
(\ref{dtot}) is a subset of $\mathbb{R}$ and represents the entire range of
allowable values for the macrovariables $\theta^{k}$. The information
geometric entropy (IGE, \cite{cafarophd,cafaroPD,cafaroChaos,cafaroAMC}) is a
proposed measure of temporal complexity of geodesic paths within the IGAC
framework and is defined as,
\begin{equation}
\mathcal{S}_{\mathcal{M}_{s}}\left(  \tau\right)  \overset{\text{def}}{=}%
\ln\widetilde{vol}\left[  \mathcal{D}_{\boldsymbol{\theta}}\left(
\tau\right)  \right]  \text{,} \label{IGE}%
\end{equation}
where the average dynamical statistical volume\textbf{\ }$\widetilde
{vol}\left[  \mathcal{D}_{\boldsymbol{\theta}}\left(  \tau\right)  \right]  $
is given by,
\begin{equation}
\widetilde{vol}\left[  \mathcal{D}_{\boldsymbol{\theta}}\left(  \tau\right)
\right]  \overset{\text{def}}{=}\frac{1}{\tau}\int_{0}^{\tau}vol\left[
\mathcal{D}_{\boldsymbol{\theta}}\left(  \tau^{\prime}\right)  \right]
d\tau^{\prime}\text{.} \label{rhs}%
\end{equation}
Observe that the tilde symbol in Eq. (\ref{rhs}) denotes the operation of
temporal average. The volume\textbf{\ }$vol\left[  \mathcal{D}%
_{\boldsymbol{\theta}}\left(  \tau^{\prime}\right)  \right]  $\textbf{\ }in
the RHS of Eq. (\ref{rhs}) is defined as,
\begin{equation}
vol\left[  \mathcal{D}_{\boldsymbol{\theta}}\left(  \tau^{\prime}\right)
\right]  \overset{\text{def}}{=}\int_{\mathcal{D}_{\boldsymbol{\theta}}\left(
\tau^{\prime}\right)  }\rho\left(  \theta^{1}\text{,..., }\theta^{n}\right)
d^{n}\boldsymbol{\theta}\text{,} \label{v}%
\end{equation}
where $\rho\left(  \theta^{1}\text{,..., }\theta^{n}\right)  $ is the
so-called Fisher density and is equal to the square root of the determinant
$g\left(  \boldsymbol{\theta}\right)  $ of the Fisher-Rao information metric
tensor $g_{\mu\nu}\left(  \boldsymbol{\theta}\right)  $,
\begin{equation}
\rho\left(  \theta^{1}\text{,..., }\theta^{n}\right)  \overset{\text{def}}%
{=}\sqrt{g\left(  \boldsymbol{\theta}\right)  }\text{.}%
\end{equation}
We remark that the expression in Eq. (\ref{v}) can be made more transparent
for statistical manifolds with information metric tensor whose determinant can
be factorized as follows,
\begin{equation}
g\left(  \boldsymbol{\theta}\right)  ={\displaystyle\prod\limits_{i=1}^{n}%
}g_{i}\left(  \theta^{i}\right)  \text{.}%
\end{equation}
Indeed, in such cases, the IGE in Eq. (\ref{IGE})\textbf{ }becomes
\begin{equation}
\mathcal{S}_{\mathcal{M}_{s}}\left(  \tau\right)  =\ln\left\{  \frac{1}{\tau
}\int_{0}^{\tau}\left[  {\displaystyle\prod\limits_{i=1}^{n}}\left(
\int_{\tau_{0}}^{\tau_{0}+\tau^{\prime}}\sqrt{g_{i}\left[  \theta^{i}\left(
\alpha\right)  \right]  }\frac{d\theta^{i}}{d\alpha}d\alpha\right)  \right]
d\tau^{\prime}\right\}  \text{.} \label{IGEmod1}%
\end{equation}
Observe that within the IGAC framework, the asymptotic behavior of
$\mathcal{S}_{\mathcal{M}_{s}}\left(  \tau\right)  $ is used to characterize
the complexity of the statistical models being analyzed. For this reason, one
usually considers the quantity $\mathcal{S}_{\mathcal{M}_{s}}^{\left(
\text{asymptotic}\right)  }\left(  \tau\right)  $,
\begin{equation}
\mathcal{S}_{\mathcal{M}_{s}}^{\left(  \text{asymptotic}\right)  }\left(
\tau\right)  \approx\lim_{\tau\rightarrow\infty}\left[  \mathcal{S}%
_{\mathcal{M}_{s}}\left(  \tau\right)  \right]  \text{,}%
\end{equation}
the leading asymptotic term in the IGE. The integration space $\mathcal{D}%
_{\theta}\left(  \tau^{\prime}\right)  $ in Eq. (\ref{v}) is given by,
\begin{equation}
\mathcal{D}_{\boldsymbol{\theta}}\left(  \tau^{\prime}\right)  \overset
{\text{def}}{=}\left\{  \boldsymbol{\theta}:\theta^{k}\left(  \tau_{0}\right)
\leq\theta^{k}\leq\theta^{k}\left(  \tau_{0}+\tau^{\prime}\right)  \right\}
\text{,} \label{is}%
\end{equation}
where $\theta^{k}=\theta^{k}\left(  \alpha\right)  $ with $\tau_{0}\leq
\alpha\leq\tau_{0}+\tau^{\prime}$ and $\tau_{0}$ denoting the initial value of
the affine parameter $\alpha$ such that {[Appendix \ref{DiffGeom}]},
\begin{equation}
\frac{d^{2}\theta^{k}\left(  \alpha\right)  }{d\alpha^{2}}+\overline{\Gamma
}_{lm}^{k}\frac{d\theta^{l}}{d\alpha}\frac{d\theta^{m}}{d\alpha}=0\text{,}
\label{ge}%
\end{equation}
{with $\overline{\Gamma}_{lm}^{k}$ denoting the Christoffel symbols of the
Riemannian connection $\overline{\nabla}$.} The integration space
$\mathcal{D}_{\boldsymbol{\theta}}\left(  \tau^{\prime}\right)  $ in Eq.
(\ref{is}) is an $n$-dimensional subspace of $\mathcal{D}_{\boldsymbol{\theta
}}^{\left(  \text{tot}\right)  }$ whose elements are $n$-dimensional
macrovariables $\left\{  \boldsymbol{\theta}\right\}  $ whose components
$\theta^{k}$ are bounded by specified limits of integration $\theta^{k}\left(
\tau_{0}\right)  $ and $\theta^{k}\left(  \tau_{0}+\tau^{\prime}\right)  $.
The integration of the $n$-dimensional set of coupled nonlinear second order
ordinary differential equations in Eq. (\ref{ge}) determines the (temporal)
functional form of such limits.

The IGE in Eq. \eqref{IGE} at a given instant is essentially the logarithm of
the volume of the effective parameter space explored by the system at that
very instant. The temporal average has been introduced in order to average out
the possibly very complex fine details of the entropic dynamical description
of the system on the curved statistical manifold. Furthermore, the long-term
asymptotic temporal behavior is considered in order to characterize in a
proper fashion the selected dynamical indicators of chaoticity, including
Lyapunov exponents and entropic quantities, by eliminating the effects of
transient effects which enter the computation of the expected value of the
volume of the effective parameter space. Therefore, the IGE is built to
provide an asymptotic coarse-grained inferential description of the complex
dynamics of a system in the presence of limited information. The construction
and interpretation of the IGE shares similarities with the logical and
thermodynamic depths \cite{B90,S88}. The logical depth \cite{B90} is
considered to be a good measure of statistical complexity \cite{L88} where the
correlated structure of the system's constituents is essential to determine
the complex path connecting the initial and final states of the system being
considered. Specifically, it is a time measure of complexity and represents
the run time needed by a universal Turing machine executing the minimal
program to reproduce a given pattern. The run time is obtained by a suitable
averaging procedure over the various programs that will accomplish the task by
weighting shorter programs more heavily. Therefore, the logical depth of any
system is defined if a suitably coarse-grained description of it is encoded
into a bit string. On the other hand, the thermodynamic depth of a process is
a structural measure of complexity and it represents the difference between
the system's coarse- and fine-grained entropy. The depth of a macrostate
reached by a particular trajectory is proportional to $-\ln p_{j}$ where
$p_{j}$ is the probability of $j$-th trajectory. The set $\left\{
p_{j}\right\}  $ represents probabilities which are consistent with all the
measurements that have been made on the system during its history.The
arbitrariness and lack of explanation of how the macrostates of the system
leading to the formation of the path-trajectory are selected \cite{Crutch99}
is an important objection to the thermodynamic depth. Within the IGE
construction, the selection of explored macrostates by the system occurs in a
manner as objective as possible since it relies on the universal MrE updating
method \cite{Giffin06}. The MrE method of determining macroscopic paths makes
no mention of randomness or other incalculable quantities. It simply chooses
the distribution (that is, the macrostate) with the maximum entropy allowed by
the information constraints. Thus, it selects the most uninformative
distribution of microstates possible. If we chose a probability distribution
with lower entropy, then we would assume information we do not possess. In
addition, to choose one with a higher entropy would violate the constraints of
the information we do possess.

The effectiveness of the IGAC theoretical approach so far described has been
proved also for particular subsets of the exponential family Eq. \eqref{exp},
namely Gaussian statistical manifolds. Using the notation used in Section II,
an $n$-dimensional multivariate Gaussian statistical model is defined in terms
of the following relations,
\begin{align}
&  \mathcal{X}\overset{\text{def}}{=}\mathbb{R}^{k},\quad n\overset
{\text{def}}{=}k+\frac{k(k+1)}{2},\quad\boldsymbol{\theta}\overset{\text{def}%
}{=}[\mu,\Sigma]\text{,}\nonumber\\
& \nonumber\\
&  \Theta\overset{\text{def}}{=}\{[\mu,\Sigma]\ |\ \mu\in\mathbb{R}%
^{k},\ \Sigma\in\mathbb{R}^{k\times k}:\ \mbox{positive definite}\}\text{,}%
\nonumber\\
& \nonumber\\
&  p(\boldsymbol{x};\boldsymbol{\theta})\overset{\text{def}}{=}\frac
{\exp\left[  -\frac{1}{2}(\boldsymbol{x}-\mu)^{\top}\ \Sigma^{-1}%
\ (\boldsymbol{x}-\mu)\right]  }{\left[  (2\pi)^{k}\det\Sigma\right]
^{1/2}\text{.}} \label{multi}%
\end{align}
In Eq. (\ref{multi}), $\mu$ is the $k$-dimensional mean vector while $\Sigma$
denotes the $k\times k$ (symmetric) covariance matrix. Specifically, in Refs.
\cite{cafaroPD,cafaroijtp08}, the IGAC is used to investigate the geometry and
the entropic dynamics of an uncorrelated Gaussian model with $l$ degrees of
freedom, each one characterized by two pieces of relevant information, its
mean and its variance. It is observed that the scalar curvature of the $2l$
dimensional statistical manifold is proportional to the number of degrees of
freedom of the system, $\mathcal{R}_{\mathcal{M}_{s}}=-l$. Similarly, from a
dynamical standpoint, it happens for the IGE in \eqref{IGE},
\begin{equation}
\mathcal{S}_{\mathcal{M}_{s}}\left(  \tau\right)  \overset{\tau\rightarrow
\infty}{\sim}l\lambda\tau\text{,}%
\end{equation}
$\label{entropy18}$where $\lambda$ denotes the maximum positive Lyapunov
exponent that characterizes the statistical model. Finally, by integrating the
geodesic deviation equations, it is found that, in the asymptotic limit, the
Jacobi vector field intensity $J_{\mathcal{M}_{s}}$ exhibits exponential
divergence and is proportional to the number of degrees of freedom $l$,
$J_{\mathcal{M}_{s}}\left(  \tau\right)  \overset{\tau\rightarrow\infty}{\sim
}l\exp\left(  \lambda\tau\right)  \text{.}$ Observing that the exponential
divergence of the Jacobi vector field intensity $J_{\mathcal{M}_{s}}$ is a
classical feature of chaos, it was argued that $\mathcal{R}_{\mathcal{M}_{s}}%
$, $\mathcal{S}_{\mathcal{M}_{s}}$ and $J_{\mathcal{M}_{s}}$ behave as proper
indicators of chaoticity and are proportional to the number of
Gaussian-distributed microstates of the system. This proportionality was the
first example in the literature of a possible substantial link among
information geometric indicators of chaoticity for probabilistic descriptions
of dynamical systems.

Further investigation of the IGAC is carried out for correlated Gaussian
statistical models in \cite{aliphysica10} to analyze the information
constrained dynamics of a system with two microscopic degrees of freedom.
These degrees of freedom are assumed to be represented by two correlated
Gaussian-distributed microvariables characterized by the same variance. The
presence of correlations at the microscopic level leads to the emergence of an
asymptotic information geometric compression of the statistical macrostates
explored by the system at a faster rate than that observed in absence of
correlations. In particular, it is found that in the asymptotic limit
\cite{aliphysica10},
\begin{equation}
\left[  \exp(\mathcal{S}_{\mathcal{M}_{s}}\left(  \tau\right)  )\right]
_{\text{{\tiny correlated}}}\overset{\tau\rightarrow\infty}{\sim}f\left(
r\right)  \, \left[  \exp(\mathcal{S}_{\mathcal{M}_{s}}\left(  \tau\right)
)\right]  _{\mbox{\tiny uncorrelated}}\text{,}%
\end{equation}
where $f\left(  r\right)  $ , with $0\leq$ $f\left(  r\right)  \leq1$, is a
monotonic decreasing compression factor for any value of the correlation
coefficient $r$ in the open interval $\left(  0\text{, }1\right)  $. This
finding represents an important and explicit connection between correlations
at the \emph{microscopic} level and complexity at the \emph{macroscopic} level
in the probabilistic description of dynamical systems within the IGAC
theoretical approach. The importance of this result is twofold: first, it
provides a clear description of the effect of information encoded in
microscopic variables on experimentally observable quantities defined in terms
of macroscopic variables; second, it neatly exhibits the behavioral change of
the macroscopic complexity of a statistical model caused by the presence of
correlations in the underlying microscopic level.

In Ref. \cite{felice15}, bivariate and trivariate Gaussian statistical models
with different correlational structures were considered and their effect on
the asymptotic behavior of the IGE \eqref{IGE} was observed. Essentially, it
was reported that the complexity of entropic inferences not only depends on
the amount of available pieces of information but also on the manner in which
such pieces are correlated. In particular, for a trivariate statistical model
with only two correlated degrees of freedom, the asymptotic temporal behavior
of the information geometric complexity ratio between the correlated and
uncorrelated cases exhibits a non monotonic behavior in terms of the
correlation parameter $r$ assuming a value equal to zero at the extrema of the
allowed range of $r$,
\begin{equation}
\frac{\left(  \exp\left[  \mathcal{S}_{\mathcal{M}_{s}}^{\text{{\tiny (mildly
connected)}}}\left(  \tau\right)  \right]  \right)  _{\text{{\tiny correlated}
}}}{\left(  \exp\left[  \mathcal{S}_{\mathcal{M}_{s}}^{\text{{\tiny (mildly
connected)}}}\left(  \tau\right)  \right]  \right)
_{\text{{\tiny uncorrelated}}}}\overset{\tau\rightarrow\infty}{\sim}%
\sqrt{\frac{3\left(  1-2r^{2}\right)  }{3-4r}}\text{.} \label{cacchio2}%
\end{equation}
However, for closed configurations (that is to say, bivariate and trivariate
models with all microscopic variables correlated to each other) the complexity
ratio exhibits a monotonic behavior in terms of the correlation parameter. For
instance, in the fully-connected trivariate Gaussian case, it was found that
\begin{equation}
\frac{\left(  \exp\left[  \mathcal{S}_{\mathcal{M}_{s}}^{\text{{\tiny (fully
connected)}}}\left(  \tau\right)  \right]  \right)  _{\text{{\tiny correlated}
}}}{\left(  \exp\left[  \mathcal{S}_{\mathcal{M}_{s}}^{\text{{\tiny (fully
connected)}}}\left(  \tau\right)  \right]  \right)
_{\text{{\tiny uncorrelated}}}}\overset{\tau\rightarrow\infty}{\sim}%
\sqrt{1+2r}\text{,} \label{cacchio}%
\end{equation}
with $0.5<r<1$. Note that a fully (mildly) connected lattice denotes a network
with a higher (lower) connectivity structure. From Eq. (\ref{cacchio}), it is
clear that in the fully connected trivariate case no peak arises and a
monotonic behavior in $r$ of the information geometric complexity ratio is
observed.\textbf{ }In the mildly connected trivariate case of Eq.
(\ref{cacchio2}), instead, a peak in the information geometric complexity
ratio is recorded at $r_{\text{peak}}=0.5$ $\geq0$. In analogy to the ideal
scenario of minimum energy spin configurations in statistical physics
\cite{mackay03,landau05}, within the IGE based approach, one would desire a
configuration of minimum complexity in order to make reliable macroscopic
predictions. Our findings in Eqs. (\ref{cacchio2}) and (\ref{cacchio}) show a
dramatically distinct behavior between the mildly connected and the fully
connected trivariate Gaussian configurations. This behavior can be ascribed to
the fact that when carrying out statistical inferences with positively
correlated Gaussian random variables, the system appears frustrated in the
fully connected case. This happens because the maximum entropy favorable
scenario seems to be incompatible with the ideal scenario of minimum
complexity. Just like certain spin configurations are not particularly
favorable from an energy standpoint, certain lattice configurations in the
presence of correlations are not especially favorable from a statistical
inference perspective of minimum complexity. In conclusion, based upon these
findings, it was argued in Ref. \cite{felice15} that the impossibility of
reaching the most favorable configuration for certain correlational structures
from an entropic inference viewpoint leads to an information geometric analog
of the frustration effect that occurs in statistical physics when loops are
present \cite{sadoc06}.

For further details on the IGAC and its applications, we refer to \cite{ali17}.


\section{Quantum outlook}

In this section, we consider the use of IG methods to characterize quantum PTs
of second-order. We also briefly discuss how to capture the concept of
complexity in quantum systems.

\subsection{ Phase transitions}

As pointed out in Section III, within the classical information geometric
setting, one of the main established findings concerns the fact that the Ricci
scalar curvature $\mathcal{R}$ is positive definite and diverges as $\xi^{d}$
at the critical point of a second order PT. Information geometric methods can
also be applied to quantum systems. Pioneering works along this line of
investigation were carried out by Provost and Vallee in Ref. \cite{PV80} where
they introduced a Riemannian metric tensor from the Hilbert space structure of
quantum states. The application of such theoretical research efforts in order
to study zero-temperature second-order quantum phase transitions is more
recent \cite{ZGC07}. {We remark that the IG of quantum PTs has different
qualitative features with respect to those of the IG of classical PTs.} For
classical systems, the geometric parameters are thermodynamic variables and
their Legendre transforms. Instead, for quantum systems, geometric parameters
are represented by quantities that appear in the Hamiltonian of the system,
for instance, the coupling constants of the theory.

\subsubsection{Curvature analysis}

Let us consider a smooth family of Hamiltonians $\mathcal{H}(\lambda)$ in a
Hilbert-space $\mathbb{H}$ where $\lambda\in\mathcal{M}$ denotes a parameter
that lives in the parameter manifold $\mathcal{M}$. Given, for the sake of
simplicity, a unique ground-state $|\Psi_{0}(\lambda)\rangle\in\mathbb{H}$ for
the Hamiltonian $\mathcal{H}(\lambda)$, we can define the map $\Psi
_{0}:\mathcal{M}\rightarrow\mathcal{H}$ in such a manner that it associates to
each parameter value the ground-state of the corresponding Hamiltonian.
Indeed, this map can be regarded as a map between $\mathcal{M}$ and the
projective space $\mathbb{PH}$. This latter space is a metric space equipped
with the Fubini-Study distance
\begin{equation}
d_{\text{FS}}(\psi,\phi)\overset{\text{def}}{=}\arccos\left[  \mathcal{F}%
(\psi,\phi)\right]  \text{,} \label{FSmetric}%
\end{equation}
where $\mathcal{F}(\psi,\phi)$ is the fidelity between the quantum pure states
$|\psi\rangle$ and $|\phi\rangle$,
\begin{equation}
\mathcal{F}(\psi,\phi)\overset{\text{def}}{=}|\langle\psi|\phi\rangle|\text{.}
\label{fidelity}%
\end{equation}
In Ref. \cite{W81}, it was noticed that $d_{\text{FS}}(\psi,\phi)$ in Eq.
(\ref{FSmetric}) is the maximum of the Fisher-Rao information metric distance
between the probability distributions that arise from $|\psi\rangle$ and
$|\phi\rangle$. Specifically, given a complete set of rank one projectors
$\{|i\rangle\langle i|\}_{i=1}^{\mathrm{{dim}\,}\mathbb{H}}$, one can consider
two probability distributions $p_{i}\overset{\text{def}}{=}|\langle
i|\psi\rangle|^{2}$ and $q_{i}\overset{\text{def}}{=}|\langle i|\phi
\rangle|^{2}$. In particular, for $p$ and $q$ infinitesimally close to each
other, the Fisher-Rao information metric distance (that represents the
distinguishability metric on the space of probability distributions) between
$p$ and $q$ becomes \cite{braunstein94},
\begin{equation}
ds_{\text{FR}}^{2}\overset{\text{def}}{=}\sum_{i=1}^{\mathrm{{dim}%
\,}\mathbb{H}}\frac{dp_{i}^{2}}{p_{i}}=\sum_{i=1}^{\mathrm{{dim}\,}\mathbb{H}%
}p_{i}(d\ln p_{i})^{2}\text{.}%
\end{equation}
The previously mentioned maximum of the Fisher-Rao information metric distance
is computed over all possible projective measurements. In addition to being a
metric space, the complex projective space $\mathbb{PH}$ is also endowed with
a Riemannian structure, that is, a metric tensor. Indeed, when the fidelity
$\mathcal{F}$ is very close to one, it can be shown that
\begin{equation}
d_{\text{FS}}^{2}(\psi,\psi+\delta\psi)\simeq2(1-\mathcal{F})\text{.}%
\end{equation}
In addition, since
\begin{equation}
\mathcal{F}(\psi,\psi+\delta\psi)\simeq|1+\langle\psi|\delta\psi
\rangle+(1/2)\langle\psi|\delta^{2}\psi\rangle|^{2}\text{,}%
\end{equation}
using this approximate expression for $\mathcal{F}$ together with the
normalization condition for the quantum state $|\psi\rangle$, one obtains
\begin{equation}
ds^{2}\overset{\text{def}}{=}d_{\text{FS}}^{2}(\psi,\psi+\delta\psi
)=\langle\delta\psi|\delta\psi\rangle-|\langle\psi|\delta\psi\rangle
|^{2}=\langle\delta\psi|(1-|\psi\rangle\langle\psi|)\delta\psi\rangle
\text{\ .} \label{FS1}%
\end{equation}
Eq. (\ref{FS1}) defines a complex metric over $\mathbb{PH}$ also known as
quantum geometric tensor \cite{PV80}. The real (imaginary) part of the quantum
geometric tensor defines a Riemannian metric tensor (a symplectic form) on
$\mathbb{PH}$. To uncover the expression of the metric in the parameter
manifold $\mathcal{M}$ induced via a pull-back by the ground state mapping
$\Psi_{0}$ introduced above, we define
\begin{equation}
\delta|\Psi_{0}(\lambda)\rangle\overset{\text{def}}{=}\sum_{\mu=1}%
^{\mathrm{{dim}\,\mathcal{M}}}|\partial_{\mu}\Psi_{0}\rangle d\lambda^{\mu
}\text{,}%
\end{equation}
where $\partial_{\mu}\overset{\text{def}}{=}\partial/\partial\lambda^{\mu}$
with $1\leq\mu\leq\mathrm{{dim}\,\mathcal{M}}$. Then, using Eq. (\ref{FS1}),
we readily obtain $ds^{2}=g_{\mu\nu}d\lambda^{\mu}d\lambda^{\nu}$ with,
\begin{equation}
g_{\mu\nu}\overset{\text{def}}{=}\operatorname{Re}\langle\partial_{\mu}%
\Psi_{0}|\partial_{\nu}\Psi_{0}\rangle-\langle\partial_{\mu}\Psi_{0}|\Psi
_{0}\rangle\langle\Psi_{0}|\partial_{\nu}\Psi_{0}\rangle\text{,}
\label{g_munu}%
\end{equation}
where $\operatorname{Re}$ denotes the real part of a complex number. Inserting
in \eqref{g_munu} the spectral resolution $I=\sum_{n}|\Psi_{n}(\lambda
)\rangle\langle\Psi_{n}(\lambda)|$ and differentiating the eigenvalue equation
$\mathcal{H}(\lambda)|\Psi_{0}(\lambda)\rangle=E_{0}(\lambda)|\Psi_{0}%
(\lambda)\rangle$, we find
\begin{equation}
g_{\mu\nu}=\operatorname{Re}\sum_{n\neq0}\frac{\langle\Psi_{0}(\lambda
)|\partial_{\mu}H|\Psi_{n}(\lambda)\rangle\langle\Psi_{n}(\lambda
)|\partial_{\nu}H|\Psi_{0}(\lambda)\rangle}{[E_{n}(\lambda)-E_{0}%
(\lambda)]^{2}}\ \text{,} \label{pert}%
\end{equation}
where $|\Psi_{n}(\lambda)\rangle$ are the eigenvectors of $\mathcal{H}%
(\lambda)$. The expression of $g_{\mu\nu}$ in Eq. (\ref{pert}) clearly
suggests that at the critical points, where one of $[E_{n}(\lambda_{c}%
)-E_{0}(\lambda_{c})]\geq0$ vanishes in the thermodynamical limit
\cite{sachdev99}, the metric tensor exhibits a singular behavior.

\paragraph{The transverse $XY$spin chain model.}

In order to explicitly see how the singularities of $g_{\mu\nu}$ arise, we
will discuss the paradigmatic case of the $XY$ model following \cite{ZGC07}.
Let us start from systems of quasi-free fermions that are defined by the
quadratic Hamiltonian
\begin{equation}
\mathcal{H}\overset{\text{def}}{=}\sum_{i,j=1}^{L}c_{i}^{\dag}A_{ij}%
c_{j}+\frac{1}{2}\sum_{i,j=1}^{L}\left(  c_{i}^{\dag}B_{ij}c_{j}^{\dag}%
+c_{j}B_{ji}c_{i}\right)  \text{,} \label{eq:Hfermions}%
\end{equation}
where $c_{i}$ ($c_{i}^{\dag}$) are annihilation (creation) operators of $L$
fermionic modes, while $A$ and $B$ are $L\times L$ real matrices symmetric and
anti-symmetric, respectively. The set of ground states for the Hamiltonian
system described in Eq. \eqref{eq:Hfermions} is parametrized by orthogonal
$L\times L$ real matrices $T$ that yield the unitary part of the polar
decomposition of the matrix $Z\overset{\text{def}}{=}A-B$. Then, one can prove
that,
\begin{equation}
\mathcal{F}(Z,Z^{\prime})\overset{\text{def}}{=}|\langle\Psi_{Z}%
|\Psi_{Z^{\prime}}\rangle|=\sqrt{|\det[(T+T^{\prime})/2]|}\text{,} \label{fzz}%
\end{equation}
where $T$ and $T^{\prime}$ denote the unitary parts of the polar
decompositions of $Z$ and $Z^{\prime}$, respectively \cite{Zanardi07}. By
expanding the expression of $\mathcal{F}(Z,Z^{\prime})$ in Eq. (\ref{fzz})
with respect to $T^{\prime}$ around $T$, it is possible to find an explicit
expression for the infinitesimal metric distance $ds^{2}$,
\begin{equation}
ds^{2}\approx2(1-\mathcal{F})=\frac{1}{8}\mathrm{tr}\left(  dK\right)
^{2}\text{, with }K\overset{\text{def}}{=}\ln T\text{.} \label{lineel}%
\end{equation}
From Eq. (\ref{lineel}), if $K=K(\lambda)$ with $\lambda\in\mathcal{M}$, one
obtains the following expression for the metric tensor induced over
$\mathcal{M}$,
\begin{equation}
g_{\mu\nu}=\frac{1}{8}\mathrm{tr}\left(  \partial_{\mu}K\partial_{\nu
}K\right)  \text{.}%
\end{equation}
For translationally invariant Hamiltonians \eqref{eq:Hfermions}, the matrix
$K$ can always be cast in the form $K=i\oplus_{k}\theta_{k}\sigma_{k}^{y}$
with $k$ being a momentum label while $i$ denotes the complex imaginary unit.
Therefore, in this working assumption, we have
\begin{equation}
g_{\mu\nu}=\frac{1}{4}\sum_{k}\frac{\partial\theta_{k}}{\partial{\lambda^{\mu
}}}\frac{\partial\theta_{k}}{\partial{\lambda^{\nu}}}\text{.}
\label{eq:gtheta}%
\end{equation}
Now, the Hamiltonian for a periodic $XY$ spin chain with an odd number of
spins $L=2M+1$ in a transverse magnetic field $h$ can be written as
\cite{ZGC07},
\begin{equation}
\mathcal{H}\overset{\text{def}}{=}\sum_{j=-M}^{M}\left[  -\frac{1+\gamma}%
{4}\sigma_{j}^{x}\sigma_{j+1}^{x}-\frac{1-\gamma}{4}\sigma_{j}^{y}\sigma
_{j+1}^{y}+\frac{h}{2}\sigma_{j}^{z}\right]  \text{,} \label{HXY}%
\end{equation}
where $\gamma$ is the anisotropy parameter in the $x$-$y$ plane and $h$ is the
magnetic field. The critical points of this model are the lines $h=\pm1$ and
the segment $|h|<1$, $\gamma=0$ \cite{sachdev99}. The Hamiltonian \eqref{HXY}
can be cast in the form \eqref{eq:Hfermions} by the Jordan-Wigner
transformation \cite{sachdev99}. Moreover, it is translationally invariant and
Eq.\eqref{eq:gtheta} can be used with $\lambda^{1}=h$, $\lambda^{2}=\gamma$
and $\theta_{k}=\arccos\left(  (\cos x_{k}-h)/\Lambda_{k}\right)  $. The
quantity $\Lambda_{k}$ denotes the single particle energy,
\begin{equation}
\Lambda_{k}\overset{\text{def}}{=}\sqrt{[\cos x_{k}-h]^{2}+\gamma^{2}\sin
^{2}x_{k}}\text{,}%
\end{equation}
where $x_{k}\overset{\text{def}}{=}2\pi k/L$ with $-M\leq k\leq M$. In the
thermodynamic limit where $L$ is very large, one replaces the discrete
variable $x_{k}$ with a continuous variable $x$ and substitutes the sum with
an integral, that is to say,
\begin{equation}
\sum_{k=1}^{M}\rightarrow\lbrack L/(2\pi)]\int_{0}^{\pi}dx\text{.}
\label{substitution}%
\end{equation}
While at critical points this substitution is not generally feasible due to
singularities in some terms of the sums in Eq. (\ref{substitution}), when away
from critical points, the resulting integrals yield analytical formulae. For
instance, for $|h|<1$, it happens that \cite{ZGC07},
\begin{equation}
g=\frac{L}{16|\gamma|}\mathrm{diag}\left(  \frac{1}{1-{h}^{2}},\frac
{1}{(1+|\gamma|)^{2}}\right)  \text{.}%
\end{equation}
We remark that it is possible to provide a closed analytic expression
(although in a less compact form \cite{ZGC07}) for $g$ in the thermodynamical
limit for $|{h}|>1$. We only point out here that for $|{h}|>1$, the
off-diagonal elements of the metric tensor are non-zero. Having the induced
metric tensor, it becomes straightforward to compute the scalar curvature
$\mathcal{R}$. One finds that for $|h|<1$,
\begin{equation}
\mathcal{R}=-\frac{16}{L}\frac{1+|\gamma|}{|\gamma|}\text{,} \label{r1}%
\end{equation}
while for $|h|>1$,
\begin{equation}
\mathcal{R}=\frac{16}{L}\frac{|{h}|+\sqrt{{h}^{2}+\gamma^{2}-1}}{\sqrt{{h}%
^{2}+\gamma^{2}-1}}\text{.} \label{r2}%
\end{equation}
From Eqs. (\ref{r1}) and (\ref{r2}), we notice that $\mathcal{R}$ diverges on
the segment $|{h}|\leq1$, $\gamma=0$, while it is only discontinuous on the
lines ${h}=\pm1$. Thus, while the components of the metric tensor on the
parameter manifold generally diverge at a quantum PT (see Eq.\ (\ref{pert})),
the same is not true for the scalar curvature.

\paragraph{The Dicke model.}

Considering the so-called Dicke model of quantum optics in the thermodynamic
limit, for finite values of detuning \cite{Dey12}, similar results as those
found in Ref. \cite{ZGC07} can be obtained. The Dicke model mimics the dipole
interaction between a single bosonic mode on a system of $N$ two-state atoms
with Hamiltonian given by \cite{gerry},
\begin{equation}
\mathcal{H}\left(  \omega\text{, }\lambda\right)  \overset{\text{def}}%
{=}\omega_{0}J_{z}+\omega a^{\dagger}a+\frac{\lambda}{\sqrt{N}}\left(
a^{\dagger}+a\right)  \left(  J_{+}+J_{-}\right)  \text{,} \label{Dicke}%
\end{equation}
where
\begin{equation}
2J_{z}\overset{\text{def}}{=}\sum_{i=1}^{N}\sigma_{z}^{(i)},\quad
\mbox{and}\quad2J_{\pm}\overset{\text{def}}{=}\sum_{i=1}^{N}\left(  \sigma
_{x}^{(i)}\pm\sigma_{y}^{(i)}\right)  ,
\end{equation}
$\omega_{0}$ is the difference between the energies of the two states of the
atom, $\omega$ is the frequency of the bosonic mode with annihilation and
creation operators $a,a^{\dag}$, respectively. {The quantity $\lambda$ denotes
the atom-field coupling strength and $N$ is the number of two-states atoms.}
In Ref. \cite{Dey12}, assuming to be in the thermodynamic limit {(i.e.
$N\rightarrow\infty$) and being in the rotating wave approximation (i.e.
$\left(  a^{\dagger}+a\right)  \left(  J_{+}+J_{-}\right)  \approx\left(
a^{\dagger}J_{-}+aJ_{+}\right)  $),} the scalar curvature of the parameter
manifold that corresponds to the Hamiltonian in Eq. (\ref{Dicke}) exhibits a
regular behavior at the phase transitions for any $\omega$ value. However, the
components of the metric tensor diverge,
\begin{equation}
\lim_{\lambda\rightarrow\lambda_{c}}\mathcal{R}\left(  \omega\text{, }%
\lambda\right)  =4\text{,} \label{noproblem}%
\end{equation}
{where $\lambda_{c}\overset{\text{def}}{=}\sqrt{\omega_{0}\omega}$.} The
divergence of the components of the metric tensor is not genuine (that is, it
is physically meaningless) and can be removed upon a suitable coordinates
transformation. Therefore, unlike the classical case where the scalar
curvature diverges at second-order critical points, in the quantum case the
geometry of the parameter manifold has to be investigated on a case-by-case
basis. The scalar curvature can be regarded as a privileged quantity to study
in the analysis of PTs because it is invariant under coordinate
transformations of the parameter space, as opposed to the metric components
(which, in turn, transform as rank two tensors). Indeed, for two-dimensional
parameter manifolds, the Riemann curvature tensor has only one independent
component, and hence leads to a unique scalar curvature. Whereas divergences
in the components of the metric tensor might be due to specific coordinate
choices (and hence can be removed by suitable coordinate transformations)
those of the scalar curvature correspond to ones that cannot be removed by
such transformations, and, therefore characterize the global (coordinate
independent) properties of the manifold.

Overall, however, the above presented results point to the fact that scalar
curvature does not seem to provide a universal characterization of
second-order quantum PTs. Hence the need to study {other geometric quantities,
like geodesic curves.}

\subsubsection{Geodesic analysis}

In addition to the notions of metric tensor and its corresponding scalar
curvature, one could employ a geodesic analysis to investigate PTs. Such a
geodesic approach was explored for classical PTs in Ref. \cite{Diosi89} and
for quantum PTs in Ref. \cite{Rez09}. The main body of work of these
investigations was based on extended numerical computations \cite{Kum12} which
were necessary due to the difficulties in integrating nonlinear sets of
coupled geodesic equations for a variety of classical and quantum models. In
summary, the main conclusion achieved was that geodesic trajectories are
confined to a single phase. This implies that on a curved parameter manifold,
points that lie in different phases are not geodesically connected. Therefore,
points on the manifold can be regarded as separated by phase transitions.
Furthermore, close to the critical point, geodesics exhibit a turnaround
behavior. This is not unexpected since the spinodal curve, being the locus of
divergences of the scalar curvature on the parameter manifold, tends to
incline the geodesics. Hence, geodesics do not cross spinodal lines. It is
well known that geodesics converge, or diverge, at singularities of the
manifold. However, the presence of geodesics that tend to converge does not
always indicate a singularity. An illustrative example is provided by a
two-sphere. In this case, great circles focus at the poles which are, however,
regular points on the sphere.

In addition to distinct curvature-based and geodesic-based analyses, phase
transitions can also be\textbf{ }analyzed by studying the convergence or
divergence of geodesic congruences near critical points. This viewpoint relies
upon the understanding of curvature effects on geodesic paths where the
evolution of a geodesic congruence is determined by means of three scalar
(i.e., coordinate-independent) parameters: expansion, shear, and rotation
parameters. In particular, the convergence of geodesics in a two-dimensional
manifold can be quantified by a so-called scalar expansion parameter $\Xi$
\cite{sayan}. To describe and understand this expansion parameter, let us
consider coordinates $x^{a}$ on the parameter manifold. Then, geodesic paths
satisfy the equation {[Appendix \ref{DiffGeom}]},
\begin{equation}
(x^{a})^{^{\prime\prime}}+\Gamma_{bc}^{a}(x^{b})^{^{\prime}}(x^{c})^{^{\prime
}}=0\text{,}%
\end{equation}
with $\Gamma_{bc}^{a}\overset{\text{def}}{=}\frac{1}{2}g^{ad}\left(
g_{db,c}+g_{dc,b}-g_{bc,d}\right)  $ denoting the Christoffel connections, and
the prime describing the derivative with respect to an affine parameter
$\lambda$ along the geodesic curves. The affine parameter is conventionally
taken as the square root of the line element, that is to say, $d\lambda
^{2}\overset{\text{def}}{=}g_{ab}dx^{a}dx^{b}$. For such an affinely
parametrized geodesic, the geodesic equations can be derived by means of a
variational principle from the Lagrangian $\mathcal{L}$,
\begin{equation}
\mathcal{L}\overset{\text{def}}{=}\frac{1}{2}g_{ab}(x^{a})^{^{\prime}}%
(x^{b})^{^{\prime}}\text{.}%
\end{equation}
Denoting the normalized tangent vectors as $u^{a}\overset{\text{def}}{=}%
(x^{a})^{^{\prime}}$, curvature effects on geodesics near criticality are
measured by the tensor
\begin{equation}
B_{\phantom{1}b}^{a}\overset{\text{def}}{=}\nabla_{b}u^{a}\text{,}%
\end{equation}
where $\nabla_{a}$ is the covariant derivative defined on a generic vector
$V^{a}$ as,
\begin{equation}
\nabla_{a}V^{b}\overset{\text{def}}{=}\partial_{a}V^{b}+\Gamma_{ac}^{b}%
V^{c}\text{.}%
\end{equation}
Finally, the expansion parameter is defined as $\Xi\overset{\text{def}}%
{=}B_{\phantom{1}a}^{a}$. In the vicinity of critical points of the parameter
manifold, $\Xi$ diverges. To compute $\Xi$, a solution for the vectors $u^{a}$
is needed and the conditions to find such solutions in an analytical manner
are discussed in Ref. \cite{tapo3}.

To recap, in a two-dimensional parameter manifold, it is possible to single
out three scalar invariants (that is, coordinate independent quantities).
These are the scalar curvature $\mathcal{R}$, the scalar expansion parameter
$\Xi$, and, finally, the line element $ds^{2}$ which is identified with an
affine parameter that measures infinitesimal distance along geodesics.

\paragraph{The transverse $XY$ spin chain model.}

As pointed out in Ref. \cite{tapo3}, there are algebraic relations among these
three scalar quantities that can reveal universal behavior in both classical
and quantum phase transitions, under the working assumption that the scalar
curvature diverges at criticality as a power law. Indeed, for both the
(classical) $1D$ Ising model and the (quantum) transverse $XY$ spin chain with
the Hamiltonian defined in Eq. (\ref{HXY}), it is possible to obtain the
following relations for the scalar curvature $\mathcal{R}$ and the scalar
expansion parameter $\Xi$ \cite{tapo3},
\begin{equation}
\mathcal{R}\sim\lambda^{-2}\text{ and, }\Xi\sim\lambda^{-1}\text{,}
\label{geoexponents}%
\end{equation}
respectively. The existence of algebraic relations among the above-mentioned
three scalar quantities leads one to suspect that there might be a generic way
to compute metrics on the parameter manifold, at least near criticality.
Indeed, in Ref. \cite{Mai15} a method in which this can be achieved by means
of ideas from scaling symmetries. Let $K^{a}$ be a homothetic vector field on
a manifold with metric $g_{ab}$ such that,
\begin{equation}
{\mathcal{L}}_{K}g_{ab}\overset{\text{def}}{=}Dg_{ab}\text{,} \label{lie}%
\end{equation}
where ${\mathcal{L}}_{K}g_{ab}$ is the Lie derivative of the metric along a
curve whose tangent is $K^{a}$ and, $D$ is a constant that can be identified
with the spatial dimension of the system which, in turn, is different from the
dimensionality of the parameter manifold. Eq. (\ref{lie}) can be rewritten
as,
\begin{equation}
K_{a;b}+K_{b;a}=Dg_{ab}\text{,} \label{homothety1}%
\end{equation}
where the semicolon denotes a covariant derivative. Eq. (\ref{homothety1}) can
be further simplified as follows,
\begin{equation}
g_{ac}K_{,b}^{c}+g_{bc}K_{,a}^{c}+g_{ab,c}K^{c}=Dg_{ab}\text{,}
\label{homothety2}%
\end{equation}
where the comma denotes an ordinary derivative with respect to the coordinate
label that follows it and repeated indices imply summation. According to the
information geometric analysis presented in Ref. \cite{Mai15} (which, in turn,
was inspired by the original work appearing in Ref. \cite{diosi84}), near
criticality, one can impose the condition that the so-called beta functions of
the theory (for further details on the beta functions, we refer to Ref.
\cite{peskin95}) are the components of a tangent vector field which is
homothetic. Then, from Eq.\eqref{homothety2}, we get a set of coupled partial
differential equations for the components of the metric. These equations, if
solvable, will lead to solutions of the metric without a detailed knowledge of
the full solution of the system. For further details, we refer to Ref.
\cite{Mai15}.

\subsection{Complexity}

From a classical physics standpoint, a complex dynamical system is
characterized by extreme sensitivity to initial conditions at fixed system
parameters. This type of complexity characterization requires use of the
concept of trajectory which, in turn, is not present in quantum physics.
Furthermore, quantum evolution is governed by a linear evolution equation
which leads to a vanishing maximal Lyapunov exponent \cite{crisanti94}, a
suitable quantifier of the sensitivity to initial conditions of the physical
system. The investigation of stability properties of both classical and
quantum evolution can be performed by means of the classical and quantum
phase-space distributions, respectively. Therefore, given the fact that both
classical and quantum mechanics can be formulated in a phase-space setting and
since it is also known that the number of Fourier components of the classical
distribution function in phase-space grows exponentially for chaotic systems
while it grows only linearly for integrable systems \cite{brumer97}, it was
proposed in Refs. \cite{benenti09, sokolov08} that the complexity of quantum
evolution can be quantified by means of the number of harmonics (that is, the
Fourier components) of the Wigner function.

Along with this statistical characterization of complexity of quantum systems,
the issue of comparing quantum and classical temporal complexity and
explaining why the former is weaker than the latter has been carried out
within the framework of IGAC. In particular, the tools of Jacobi Levi-Civita
(JLC) vector field intensity and Information Geometric Entropy (IGE) are
applied to a three-dimensional uncorrelated Gaussian model and a
two-dimensional Gaussian statistical model obtained from the higher
dimensional model via introduction of an additional information constraint
that resembles the canonical minimum uncertainty relation in quantum theory
\cite{cafaroosid}.

In view of the description given in Eq. \eqref{multi}, a three dimensional
Gaussian statistical model is characterized by the following pdf
\begin{equation}
p(x,y|\sigma_{x},\sigma_{y},\mu_{x})=\frac{1}{2\pi\sigma_{x}\sigma_{y}}%
\exp\left[  -\frac{1}{2\sigma_{x}^{2}}(x-\mu_{x})^{2}-\frac{1}{2\sigma_{y}%
^{2}}y^{2}\right]  , \label{3Gauss}%
\end{equation}
where $\sigma_{x},\sigma_{y}\in\mathbb{R}^{+}$ and $\mu_{x}\in\mathbb{R}$. At
this point, we can introduce the following macroscopic constraint,
\begin{equation}
\sigma_{x}\sigma_{y}=\chi^{2}, \label{costraintGaussian}%
\end{equation}
where $\chi^{2}$ is a positive \textit{real} constant. Therefore, by setting
$\sigma\equiv\sigma_{x}$, the two dimensional Gaussian model arising from the
one in Eq. \eqref{3Gauss} is described by the following pdf,
\begin{equation}
p(x,y|\sigma,\mu_{x})=\frac{1}{2\pi\Sigma^{2}}\exp\left[  -\frac{1}%
{2\sigma^{2}}(x-\mu_{x})^{2}-\frac{\sigma^{2}}{2\chi^{4}}y^{2}\right]  .
\label{2Gauss}%
\end{equation}
The macroscopic constraint \eqref{costraintGaussian} resembles the quantum
mechanical canonical minimum uncertainty relation where $x$ denotes the
position of a particle and $y$ its conjugate momentum.

By applying the IGE of Eq. \eqref{IGEmod1} to the Gaussian model described
above, one obtains the following relation \cite{cafaroosid},
\begin{equation}
\mathcal{S}^{\mbox{\tiny 2Dmodel}}(\tau)\overset{\tau\gg1}{\approx}\frac
{1}{\sqrt{2}}\ \mathcal{S}^{\mbox{\tiny 3Dmodel}}(\tau) \label{IGEGauss}%
\end{equation}
Eq. \eqref{IGEGauss} quantitatively shows that the IGE is softened when
approaching the two dimensional case from the three dimensional one via the
introduction of the macroscopic constraint \eqref{costraintGaussian} that is
reminiscent of Heisenberg's minimum uncertainty relation.

A further investigation that concerns the attenuation of the asymptotic
temporal growth of complexity indicators of entropic motion has been carried
out in Ref. \cite{cafaroosid}. In particular, a JLC vector field intensity
analysis was performed. A JLC vector field $J=(J^{1},\ldots,J^{n})$ is a
solution of the JLC equation [Appendix \ref{DiffGeom}],
\begin{equation}
\frac{\nabla^{2}J^{k}}{d\tau^{2}}+R_{ijr}^{k}\frac{d\theta^{i}}{d\tau}%
J^{j}\frac{d\theta^{r}}{d\tau}=0, \label{JLC}%
\end{equation}
where $R_{ijr}^{k}$ are the components of the Riemann curvature tensor and
$\theta^{i}(\tau)$ are the components of the geodesic flow upon which the IGAC
relies. Eq. \eqref{JLC} quantifies the connection between the geodesic spread
and the curvature of the given Riemannian metric. In particular, it is often
employed as an indicator of the stability/instability of the dynamics by
determining the evolution of perturbations of a given trajectory. Indeed, the
JLC vector field accounts for such a perturbation from the geodesic dynamics.
By computing JLC vector fields for the previous three and two dimensional
Gaussian models\textbf{ }in the Eqs. \eqref{3Gauss} and \eqref{2Gauss}, we
arrive at the following asymptotic relation
\begin{equation}
J^{\mbox{\tiny 2D}}(\tau)\overset{\tau\gg1}{\approx}e^{-(\Lambda_{3D}%
-\Lambda_{2D})\tau}\ J^{\mbox{\tiny 3D}}(\tau),\quad\mbox{with}\quad
\Lambda_{3D}-\Lambda_{2D}>0. \label{JLCrelation}%
\end{equation}
Eq. \eqref{JLCrelation} is quite informative since it shows that the Jacobi
vector field intensity is softened when approaching the two-dimensional case
from the three-dimensional case via the introduction of the quantum-like
macroscopic constraint \eqref{costraintGaussian}. The approach so far
described and implemented in \cite{cafaroosid} for tackling the issue of
complexity of quantum systems relies on the similarity between the information
constraint on the variances and the phase-space coarse-graining imposed by the
Heisenberg uncertainty relations suggesting that it provides a possible way of
explaining the phenomenon of suppression of classical chaos operated on by
quantization constraints.

In addition to such investigations, the complexity of quantum systems has
recently been studied in terms of less conventional techniques. One approach
relies on defining a complexity measure for a pdf and then apply it to the pdf
of certain variables of the system, principally the configuration or momentum.
Among the measures of statistical complexity applied to quantum systems, the
Fisher-Shannon measure plays a key role \cite{Sanudo08}. Indeed, the
statistical complexity of a probability distribution function constructed from
the measurement of quantum observables forming a continuous manifold has been
proposed as a suitable way to characterize the non-integrable behavior of
quantum systems. In{ the Hilbert space $L_{2}(\mathbb{R})$, a continuos
manifold of observables is defined by the operators,
\begin{equation}
O_{\omega}\overset{\text{def}}{=}Q\cos\omega-P\sin\omega,\quad\omega\in
\lbrack0,\pi\lbrack
\end{equation}
where $(Q,P)$ is a pair of }canonically conjugate observables,{ namely
position and momentum, respectively. By denoting $|o_{\omega}\rangle$ the
eigenvectors of the operator $O_{\omega}$ with eigenvalue $o_{\omega}$, the
probability distribution for the output of the measurement of $O_{\omega}$ is
defined as $\rho(o_{\omega})=|\langle o_{\omega}|\psi\rangle|^{2}$ for any
state $|\psi\rangle$ of the quantum system. Therefore the statistical
complexity in the space determined by the observable $O_{\omega}$ is defined
in terms of the product of two factors \cite{Dehesa09},
\begin{equation}
C_{FS}\overset{\text{def}}{=}I[\rho(o_{\omega})]\times J[\rho(o_{\omega})],
\label{FScomplexity}%
\end{equation}
where
\begin{align}
&  I[\rho(o_{\omega})]\overset{\text{def}}{=}\int_{-\infty}^{\infty}%
\rho(o_{\omega})\left[  \frac{d}{do_{\omega}}\ln(\rho(o_{\omega}))\right]
do_{\omega},\label{F-Scomponent}\\
&  J[\rho(o_{\omega})]\overset{\text{def}}{=}\frac{1}{2\pi e}\exp\left(
2H_{\mathrm{Shannon}}[\rho(o_{\omega})]\right)  ,
\end{align}
are the Fisher information and the entropic power of Shannon entropy,
respectively. The Shannon entropy is defined as $H_{\mathrm{Shannon}}%
[\rho(o_{\omega})]\overset{\text{def}}{=}-\int_{-\infty}^{\infty}%
\rho(o_{\omega})\ln(\rho(o_{\omega}))do_{\omega}$. }The Fisher-Shannon measure
is composed of the product of a measure of the spreading of the function and a
measure of possible oscillations thereof.{ By analyzing the complexity
\eqref{FScomplexity} it is evident that it takes different values depending on
the (phase space) parameter $\omega$.}

In \cite{manzano12}, a way to overcome this difficulty by considering the
entire phase space was proposed. To this end, two complexity measures were
defined. First, the global Fisher Shannon (GFS) complexity was considered
\begin{equation}
C_{GFS}[|\psi\rangle]\overset{\text{def}}{=}\frac{1}{\pi}\int_{0}^{\pi}%
I[\rho(o_{\omega})]\times J[\rho(o_{\omega})]d\omega. \label{GFS}%
\end{equation}
Second, representing a possibility for generating a base independent measure
of complexity, the minimum value of the usual Fisher Shannon complexity in the
complete range of $\omega$ was taken into consideration,
\begin{equation}
C_{MFS}[|\psi\rangle]\overset{\text{def}}{=}\min_{\omega\in\lbrack0,\pi
]}I[\rho(o_{\omega})]\times J[\rho(o_{\omega})]. \label{MFS}%
\end{equation}
The latter measure relies on Kolmogorov's complexity idea that the complexity
of a system must be calculated in its simplest description. The measure of Eq.
\eqref{MFS} is then called the \textit{minimum Fisher Shannon} (MFS) measure
of complexity.

For both measures, the global one and the minimum one, if on one side they
turn out to be independent of the conjugate states, on the other side they
will require changing the condition for statistical measures of complexity by
a more general condition. Relying on the fact that \textit{both, the GFS and
the MFS measures give the minimum value for the states that can be represented
as a Gaussian distribution for any $o_{\omega}$} \cite{manzano12}, the new
definition of complexity is given in terms of non-Gaussianity, that in turn is
compatible with the original Fisher Shannon complexity measure.

The quantification of complexity of motion of physical systems in terms of
probability distribution functions constructed from relevant observables of
the system, paved the way to apply information geometric methods to quantify
complexity of quantum systems. The complexity of quantum energy levels
statistics was investigated in Refs. \cite{carloMPLB,cafaroPA}, the complexity
of motion in scattering induced quantum entanglement appeared in Refs.
\cite{kim2,kim1}, while the milder nature of quantum chaoticity with respect
to classical chaos was preliminarily addressed from an information geometric
standpoint in Refs. \cite{cafaroaip,giffinaip,cafaroosid,softchaos}. The use
of differential geometric methods to investigate some aspects of complexity of
quantum systems has been very recently presented in Ref. \cite{susskind2017}.
Here, it was shown that the growth of computational complexity of a quantum
system described by a set of strongly coupled Hamiltonians exhibits similar
features as those that specify classical geodesics on a compact
two-dimensional geometry of uniform negative curvature.

Unfortunately, most of these results are not conclusive. The challenge of
uncovering a unifying measure of complexity and the very essence of the
concept of complexity itself still remains to be uncovered in full detail.
Despite substantial efforts during the last two decades or so
\cite{petz96,petz99,petz02,gibilisco03,gibilisco05,gibilisco06,gibilisco07,facchi10,ciaglia17}
, methods of information geometry extended to the quantum physical settings
have not yet reached their full maturity with regard to the quantification of
the concept of complexity of quantum evolution. In particular, to the best of
our knowledge, there is no information geometric characterization of the
concepts of structural complexity of quantum networks \cite{siomau16} nor of
quantum statistical complexity \cite{gu14}.


\section{Conclusion}

In this manuscript, we reviewed the application of IG methods to describe the
notion of complexity in both classical and, when possible, quantum physical
settings. We started from classical phase transitions since they are
intimately connected to changes in the complexity of physical systems. Then,
motivated by the observation of emergent phenomena, we considered a potential
measure of complexity that links heterogeneous systems to homogeneous ones.
This measure is the so-called Riemann geometric entropy in Eq. (\ref{entropy}%
). We then discussed an additional complexity measure that is very
well-grounded on IG methods. This measure appears in Eq. (\ref{AyComplMeas})
and is based upon the notion of stochastic interaction. Subsequently we
reported on chaoticity and complexity properties addressed within the area of
entropic motion. Within the entropic framework, the concept of information
geometric entropy in Eq. (\ref{IGEmod1}) was proposed. Finally, we reviewed IG
methods applied to the quantum setting where both features of PTs and
complexity were addressed.

\subsection{Phase transitions}

We discussed the use of information geometric methods to study second order
phase transitions in classical systems by analyzing the divergence and/or
scaling behavior of the scalar Ricci curvature in the proximity of a critical
point. We recall that the Ricci scalar curvature is a geometric invariant
quantity. Specifically, we described the IG\ characterization of PTs in four
models: i) { the one-dimensional Ising model \cite{jany89} reporting the
scaling behavior of the scalar curvature in the proximity of the critical
region in Eq. (\ref{ising1d});} ii) the one-dimensional $q$-state Potts model
\cite{dolan02,Dol98} with the scaling behavior of the Ricci scalar curvature
at the critical point given in Eq. (\ref{potts}); iii) the Ising model on
planar random graphs \cite{janke02} with scaling behavior of the scalar
curvature reported in\ Eq. (\ref{yoyo1}); iv) {the uniform random graph model
reporting the Erd\"{o}s-R\'{e}nyi phase transition in terms of a numerical
estimation \cite{franzosi16} of the Riemannian geometric entropy presented in
Eq. (\ref{nicco}).} For the first two models, the probability distributions
that define the parameter manifolds whose curvature properties are
investigated are parametrized by two parameters, the inverse temperature
$\beta$ and the external magnetic field intensity $h$, respectively.
Furthermore, both global and local properties of phase transitions described
in terms of the scalar curvature of the parameter manifold and the components
of the metric tensor, respectively, were considered for such models. For the
third model, we only discussed the scaling behavior of the scalar curvature.
Finally, in the fourth model, we considered a Riemannian geometric entropy,
constructed from the volume of statistical manifold associated to a network
(see Eq. \eqref{entropy}), which encodes both structural and statistical
complexity aspects. We emphasize that for classical systems, including fluid
and spin systems, the curvature always exhibits a divergence at the critical
point of a second-order phase transition. Moreover, in the classical scenario,
there is a fairly clear connection between the correlation length $\xi$ and
the scalar curvature $\mathcal{R}$ as reported in\ Eq. (\ref{achi}).

\subsection{Complexity}

We reviewed three approaches to complexity characterization using IG methods.
First, we discussed a Riemannian geometric entropy measure of complexity as
reported in Eq. (\ref{entropy}) for various complex networks. By associating
smooth systems (Riemannian manifolds) to discrete systems (network), this
approach allowed to transfer the use of well established geometric tools to
the setting of discrete mathematics. In particular, the detection of the
Erd\"{o}s-R\'{e}nyi phase transition suggested to consider such a Riemannian
geometric entropy as a good candidate to measure the degree of organization of
networks of complex systems. Indeed, one of the strengths of this Riemannian
geometric entropy is that it allows to quantify both structural and
statistical features of complex networks \cite{franzosi16}. Second, exploiting
the information geometric concept of hierarchical structure of probability
distributions, we discussed the complexity of a physical system in terms of
the interactions among its interacting units at different scales of
description. One of the main advantages of the complexity measure in Eq.
(\ref{AyComplMeas}) is that it takes into account the decomposition of a
complex systems in terms of an interaction hierarchy \cite{Aypre}. Third,
moving to a dynamical scenario, we described methods of information geometry
combined with inductive inference techniques in order to quantify the
complexity of entropic motion on curved statistical manifolds underlying a
probabilistic description of physical systems in the presence of partial
knowledge. This approach uses essentially three information geometric
indicators of complexity: curvature, Jacobi vector field intensity, and the
so-called information geometric entropy as defined in Eq. (\ref{IGE}). The
main advantage of this information geometric approach consists in providing a
flexible platform for making statistical predictions in the presence of
limited information about complex systems of arbitrary nature, in principle.
At the same time, one of the essential limitations of this theoretical scheme
is the lack of a corresponding fully developed quantum structure for
quantifying the complexity of quantum mechanical systems \cite{ali17}. We
point out that the concepts of volumes of curved parameter manifolds play a
key role in the first and third information geometric approaches to complexity
characterizations. Instead, the Kullback-Leibler divergence $D\left(
p||q\right)  $ and the distance to an exponential family $D\left(
p||\mathcal{E}\right)  $ in Eq. (\ref{expdist}) are the essential elements
needed to introduce a notion of complexity that relies upon the nature of
interactions among the interacting units of the physical systems at different
scales of description.

\subsection{Quantum framework}

Extending our analysis to the quantum setting, we reviewed the use of
information geometric techniques to investigate second order phase transitions
by analyzing the singular behavior of the metric tensor at the critical points
in the thermodynamical limit. In the quantum setting, the metric structure is
defined on parameter manifolds whose points are parameterized by means of the
coupling constants that appear in the Hamiltonian. More specifically, we
considered the information geometric characterization of phase transitions for
the periodic $XY$ spin chain with an odd number of spins in a transverse
magnetic field. Both a curvature-analysis (see Eqs. (\ref{r1}) and (\ref{r2})
together with Ref. \cite{ZGC07}) and a geodesic-analysis (for more details,
see Ref. \cite{Kum12}) were considered for this specific model. The geometric
parameters used in this case were the anisotropy parameter $\gamma$ and the
magnetic field intensity $h$. For this system the components of the metric
tensor on the parameter manifold diverge at the quantum phase transition,
while the Ricci scalar curvature does not (see Eqs. (\ref{r1}) and (\ref{r2})
for the scaling behavior of the curvature). We point out that, in general, the
information geometric analysis of phase transitions in the quantum setting
requires more caution since statements are based upon the divergence of metric
components which are not coordinate independent quantities and, therefore,
singularities may be removable and unphysical. Such a scenario does not appear
in the classical setting where the geometric invariance of the Ricci scalar
curvature guarantees statements based upon the global properties of the
parameter manifold. For example, the information geometric analysis (for more
details, see Eq. \ref{noproblem} and Ref. \cite{Dey12}) of the Dicke model of
quantum optics in the thermodynamic limit for finite values of detuning
confirms this type of behavior. We pointed out that in addition to distinct
curvature-based and geodesic-based analyses, phase transitions can be analyzed
in terms of the divergence of geodesic congruences in the proximity of
critical points. This line of investigation relies upon the fact that
curvature affects geodesic paths and, in addition, the evolution of a geodesic
congruence is determined by means of three scalar parameters \cite{tapo3}:
expansion, shear, and rotation parameters. The scaling behavior of both the
scalar curvature and the expansion for the (classical) $1D$ Ising model and
the (quantum) transverse $XY$ spin chain are reported in Eq.
(\ref{geoexponents}). For further interesting descriptions of criticality and
quantum phase transitions in spin chains by means of the Fisher information,
we refer to Refs. \cite{marzolino13,marzolino14,marzolino17}.

Finally, despite the absence of a fully-developed quantum information
geometric approach to complexity characterization of any kind, we briefly
reported on attempts of such an endeavor. In particular, we mentioned the
softening of classical chaos operated on by quantization constraints
\cite{cafaroosid}, the information geometric complexity of statistical models
arising in the context of quantum energy levels statistics \cite{carloMPLB}
and, finally, a statistical measure of complexity for a continuous manifold of
quantum observables \cite{manzano12}.

\subsection{Concluding remarks}

From the overview presented herein several interesting questions emerge. They
include among others, the issue of how to develop an IG characterization of
time-dependent networks in which, for instance, the weights of the edges are
not kept fixed in time \cite{franzosi16}. Furthermore, there is no available
IG\ characterization of the complexity of macroscopic predictions in the
presence of limited dynamical information constraints on complex physical
systems \cite{cafaropre}. Finally, another open problem concerns the IG
complexity characterization of a convex combination of two probability
distributions and, in particular, the issue of determining whether or not the
complexities of the individual constituents are related to the complexity of
the convex combination \cite{Aypre}. We are confident that the work reported
in this manuscript can offer a valid starting point from which to pursue
serious attempts at addressing the above-mentioned open questions. At the same
time, we emphasize that the aim of this review is not that of proposing an
information geometric measure of complexity with universal applicability.
However, we do have reason to believe that the significance of the selected
findings reported here will drive more researchers to consider the possibility
of seeking for such a measure within the powerful framework of information
geometry. In particular, we point out that, to the best of our knowledge,
there does not exist any general theoretical information geometric platform
for quantifying and, to a certain extent, understanding the concept of
complexity of quantum dynamical motion and its connection to phase transitions.


\section*{Acknowledgements}

C. C. acknowledges the hospitality of the United States Air Force Research
Laboratory in Rome (New York) where part of his contribution to this work was
completed. Finally, the authors thank the referees for constructive criticism
leading to an improved version of this manuscript.

\pagebreak


\appendix

\section{Elements of differential geometry}

\label{DiffGeom}

{In what follows, we present the main differential geometric concepts of
relevance to our work. For a more detailed description we refer to Refs.
\cite{lee97,frankel97}.}

An $n$-dimensional $C^{\infty}$ manifold $\mathcal{M}$ is a Hausdorff space
with a countable basis which is endowed with a differentiable structure,
\begin{equation}
\mathcal{U}=\left(  U_{i},\varphi_{i}\right)  _{i\in I}\text{,}
\label{DiffStruct}%
\end{equation}
where $U_{i}\subset\mathcal{M}$ is an open set and $\varphi_{i}:U_{i}%
\rightarrow\varphi(U_{i})(\subset\mathbb{R}^{n})$ is a homeomorphism,
satisfying the following conditions:

\begin{enumerate}
\item[i)] $\bigcup_{i}U_{i}=\mathcal{M}$;

\item[ii)] For any $i,j\in I$, the mapping $\varphi_{i}\circ\varphi_{j}%
^{-1}:\varphi_{j}(U_{j})\rightarrow\varphi_{i}(U_{i})$ is a $C^{\infty
}(\mathbb{R}^{n})$-function wherever it is well-defined;

\item[ii)] If $V\subset\mathcal{M}$ is an open set, $\varphi:V\rightarrow
\varphi(V)$ is a homeomorphism, $\varphi\circ\varphi_{i}^{-1}$ and
$\varphi_{i}\circ\varphi^{-1}$ are $C^{\infty}$ wherever they are
well-defined, then $(V,\varphi)\in\mathcal{U}$.
\end{enumerate}

The condition ii) is expressed as $\varphi_{i}$ and $\varphi_{j}$ being
compatible. In very simple cases, $\mathcal{M}$ is itself homeomorphic to an
open set of $\mathbb{R}^{n}$ and the differentiable structure is simply given
by $(\mathcal{M},\varphi)$ and all sets $(U_{i},\varphi_{i})$ such that
$U_{i}$ is an open set and $\varphi_{i}\circ\varphi^{-1}$ is a diffeomorphism.
The sets $U_{i}$ are called \emph{coordinate neighborhoods} and $\varphi_{i}$
coordinates. The pair $(U_{i},\varphi_{i})$ is called a \emph{local coordinate
system}. The differentiable structure allows a natural way for defining
a\emph{ differentiable function}. Let $f:\mathcal{M}\rightarrow\mathbb{R}$ be
a function, we say that it is in $C^{\infty}(\mathcal{M})$ if the mapping
$f\circ\varphi_{i}^{-1}:\varphi(U_{i})\rightarrow\mathbb{R}$ is a $C^{\infty}%
$-function for all $i\in I$. The same applies to any subset of $\mathcal{M}$
that inherits in a natural way the differentiable structure of $\mathcal{M}$.
For $p\in\mathcal{M}$, we denote with $C^{\infty}(p)$ the set of functions
whose restriction to some open neighborhood $U$ of $p$ is in $C^{\infty}(U)$.
Therefore, we identify $f$ and $g$ $\in C^{\infty}(p)$ if their restrictions
to some open neighborhood of $p$ are identical. \ We define the \emph{tangent
space} $T_{p}\mathcal{M}$ to $\mathcal{M}$ at $p$ as the set of all maps
$X_{p}:C^{\infty}(p)\rightarrow\mathbb{R}$ satisfying the following conditions:

\begin{enumerate}
\item[i)] $X_{p}(\alpha f+\beta g)=\alpha X_{p}(f)+\beta X_{p}(g)$, with
$\alpha,\beta\in\mathbb{R}$;

\item[ii)] $X_{p}(fg)=X_{p}(f)g+fX_{p}(g)$ with $f,g\in C^{\infty}(p)$;
\end{enumerate}

The quantity $X_{p}$ is called a \textit{tangent vector} and $T_{p}%
\mathcal{M}$ is in an obvious way a vector space with dimension $\dim\left[
T_{p}\mathcal{M}\right]  =n$. For each particular choice of a coordinate
system, there is a corresponding canonical basis for $T_{p}\mathcal{M}$, with
basis vectors being,
\begin{equation}
\left(  \partial_{i}\right)  _{p}(f)=\frac{\partial}{\partial x^{i}}%
f(\varphi^{-1}(x))|_{x=\varphi(p)}\text{.} \label{tangentvectorbasis}%
\end{equation}
A \textit{vector field} is a differentiable function $X:\mathcal{M}%
\rightarrow\bigcup_{p}T_{p}\mathcal{M}$ such that its composition with respect
to the projection $\pi:\bigcup_{p}T_{p}\mathcal{M}\rightarrow\mathcal{M}$
gives the\textbf{ }identity map,
\begin{equation}
\left(  \pi\circ X\right)  (p)=\pi(X_{p})=p\text{.}%
\end{equation}
The vector fields on $\mathcal{M}$ are denoted as $\mathcal{T}(\mathcal{M})$.
A \emph{Riemannian metric} $g$ is a positive symmetric tensor of rank two,
\begin{equation}
g(X,X)\geq0\text{ and, }g(X,Y)=g(Y,X).
\end{equation}
Since tensors are point-wise, we can think of a metric $g_{p}$ on each of the
tangent space $T_{p}\mathcal{M}$. A $C^{\infty}$ manifold with a positive
symmetric tensor is called a \emph{Riemannian manifold} $(\mathcal{M},g)$. A
curve $\gamma$ in $\mathcal{M}$ is a $C^{\infty}$ map $\gamma:[\alpha
,\beta]\subset\mathbb{R}\rightarrow\mathcal{M}$. It is worth noticing that a
curve is more than the sets of points in it. It involves the parametrization
and is, thus, not a purely geometric object. On the contrary, the length
$|\gamma|$ of the curve $\gamma$ is a geometric object. It is defined through
the vector field $\dot{\gamma}$,
\begin{equation}
\dot{\gamma}_{\gamma(t)}(f)=\frac{\partial}{\partial t}f(\gamma(t))\text{,
}\forall\ t\in\lbrack\alpha,\beta]\text{,}\
\end{equation}
with $f\in C^{\infty}(\mathcal{M})$ as,%
\begin{equation}
|\gamma|\overset{\text{def}}{=}\int_{\alpha}^{\beta}\ \ \sqrt{g(\dot{\gamma
},\dot{\gamma})_{\gamma(t)}}dt\text{.}%
\end{equation}
In order to compare tangent vectors $X_{p}$ and $X_{q}$ such that $p,q\in U$,
we introduce the\textbf{ }notion of an \emph{affine connection} on the
manifold $\mathcal{M}$. It is defined as an operator $\nabla$,%
\begin{equation}
\nabla:\mathcal{T}(\mathcal{M})\times\mathcal{T}(\mathcal{M})\rightarrow
\mathcal{T}(\mathcal{M})\text{,}%
\end{equation}
satisfying the following relations:

\begin{itemize}
\item[i)] $\nabla_{X}Y(\alpha Y+\beta Z)=\alpha\nabla_{X}Y+\beta\nabla_{X}Z,$
with $\alpha,\beta\in\mathbb{R}$;

\item[ii)] $\nabla_{X}(fY)=X(f)Y+f\nabla_{X}Y$, $\forall\ f\in C^{\infty
}(\mathcal{M})$;

\item[iii)] $\nabla_{fX+gY}Z=f\nabla_{X}Z+g\nabla_{Y}Z$, $\forall\ f,g\in
C^{\infty}(\mathcal{M})$.
\end{itemize}

An affine connection can be thought of as a directional derivative of vector
fields. In particular, $\nabla_{X}Y$ is the change of the vector field $Y$ in
the direction of the vector field $X$. However, an affine connection can be
defined in many ways since that change is ill defined without giving a rule
for comparing vectors in two distinct tangent spaces $T_{p}\mathcal{M}$ and
$T_{q}\mathcal{M}$. Such a rule is established by the notion of \emph{parallel
transport}. First, we say that a vector field $X$ is parallel along the curve
$\gamma:[\alpha,\beta]\rightarrow\mathcal{M}$ if,
\begin{equation}
\nabla_{\dot{\gamma}}X=0\text{,}%
\end{equation}
where $\dot{\gamma}$ denotes any vector field that represents $\frac{\partial
}{\partial t}$. Now, for any vector $X_{\gamma(\alpha)}\in T_{\gamma(\alpha
)}\mathcal{M}$, there is a unique curve of vectors,
\begin{equation}
X_{\gamma(t)}\text{, with}\ t\in\lbrack\alpha,\beta]\text{ and }X_{\gamma
(t)}\in T_{\gamma(t)}\mathcal{M}\text{,}%
\end{equation}
such that $\nabla_{\dot{\gamma}}X=0$. We then write,
\begin{equation}
X_{\gamma(\beta)}=\Pi_{\gamma}\left(  X_{\gamma(\alpha)}\right)  \text{,}%
\end{equation}
and say that $\Pi_{\gamma}$ defines parallel transport along $\gamma$. An
affine connection can be specified by choosing a local basis $\left\{
\partial_{i}\right\}  $ with $1\leq i\leq n$ for the vector-fields and by
defining the connections symbols $\Gamma_{ij}^{k}$ with $1\leq i,j,k\leq n$
as
\begin{equation}
\nabla_{\partial_{i}}\partial_{j}\overset{\text{def}}{=}\Gamma_{ij}%
^{k}\partial_{k}\text{,}%
\end{equation}
where we adopted Einstein's summation convention according to which whenever
an index appears in an expression as upper and lower index, we sum over that
index. A \emph{geodesic} is a curve with parallel tangent vector field,
\begin{equation}
\nabla_{\dot{\gamma}}\dot{\gamma}=0\ \mbox{on}\ \gamma\text{.}%
\end{equation}
Associated with the notion of a geodesic is the \emph{exponential map} induced
by the connection. For all $p\in\mathcal{M}$, $X_{p}\in T_{p}\mathcal{M}$
there is a unique geodesic $\gamma_{X_{p}}$ such that,
\begin{equation}
\gamma_{X_{p}}(0)=p\text{, and }\dot{\gamma}_{X_{p}}(0)=X_{p}\text{.}
\label{initialcond}%
\end{equation}
The geodesic is determined by the following differential equations,
\begin{equation}
\ddot{\gamma}^{k}(t)+\Gamma_{ij}^{k}\dot{\gamma}^{i}(t)\dot{\gamma}^{j}(t)=0
\label{geodesic}%
\end{equation}
together with the initial conditions \eqref{initialcond}; here $\gamma_{X_{p}%
}=(\gamma^{1}(t),\ldots,\gamma^{n}(t))$ in\textbf{ }local coordinates. Hence,
by defining for $X_{p}\in T_{p}\mathcal{M}$,
\begin{equation}
\exp(X_{p})\overset{\text{def}}{=}\gamma_{X_{p}}(1)\text{,}%
\end{equation}
we have $\exp(tX_{p})=\gamma_{X_{p}}(t)$. The exponential map is in general
well-defined at least in a neighborhood of zero in $T_{p}\mathcal{M}$ and,
moreover, can be globally defined in special cases. In general, geodesics do
not have properties of \textit{minimizing} curve length. However, on any
Riemannian manifold there is a unique affine connection $\nabla$ satisfying
the following relations,

\begin{itemize}
\item[i)] $\nabla_{X}Y-\nabla_{Y}X-[X,Y]=0$;

\item[ii)] $Xg(Y,Z)=g(\nabla_{X}Y,Z)+g(Y,\nabla_{X}Z)$;
\end{itemize}

The quantity $[X,Y]$ denotes the Lie bracket and is defined as follows,%
\begin{equation}
\lbrack X,Y](f)=X(Yf)-Y(Xf),\ \forall f\in C^{\infty}(\mathcal{M})\text{.}%
\end{equation}
This connection is called the Riemannian connection or the Levi-Civita
connection. Property i) is called \emph{torsion freeness} and property ii)
means that the parallel transport is isometric. If the connection $\nabla$ is
Riemannian, its geodesics will locally minimize curve length. When the
manifold is equipped with a Riemannian metric, it is often convenient to
specify the symbols $\Gamma_{ij,k}$,
\begin{equation}
\Gamma_{ij,k}=g\left(  \nabla_{\partial_{i}}\partial_{j},\partial_{k}\right)
\text{.}%
\end{equation}
Defining the matrix of the metric tensor and its inverse as,%
\begin{equation}
g_{ij}=g(\partial_{i},\partial_{j})\text{ and, }g^{ij}=(g_{ij})^{-1}\text{,}%
\end{equation}
respectively, the symbols $\Gamma_{ij}^{k}$ are related to those previously
defined as $\Gamma_{ij,l}$ in the following manner,
\begin{equation}
\Gamma_{ij}^{k}=g^{kl}\Gamma_{ijl}\text{.}%
\end{equation}
The Riemannian connection $\Gamma_{ij,l}$ is explicitly defined as,
\begin{equation}
\Gamma_{ij,k}\overset{\text{def}}{=}\frac{1}{2}\left(  \partial_{i}%
g_{jk}+\partial_{j}g_{ik}-\partial_{k}g_{ij}\right)  \text{.}%
\end{equation}
An important tensor field associated with a space with an affine connection is
the \emph{curvature field},
\begin{equation}
\mathcal{R}:\mathcal{T}(\mathcal{M})\times\mathcal{T}(\mathcal{M}%
)\times\mathcal{T}(\mathcal{M})\rightarrow\mathcal{T}(\mathcal{M})\text{,}%
\end{equation}
such that,%
\begin{equation}
\mathcal{R}(X,Y)Z\overset{\text{def}}{=}\nabla_{X}\nabla_{Y}Z-\nabla_{Y}%
\nabla_{X}Z-\nabla_{\lbrack X,Y]}Z\text{.}%
\end{equation}
On a Riemannian manifold $\mathcal{M}$, we also define the \emph{curvature
tensor} $\widetilde{\mathcal{R}}$ as follows,
\begin{equation}
\widetilde{\mathcal{R}}(X,Y,Z,W)\overset{\text{def}}{=}g\left(  \mathcal{R}%
(X,Y)Z,W\right)  \text{.} \label{RCurv}%
\end{equation}
The Riemann curvature tensor $\widetilde{\mathcal{R}}$ satisfies the following properties,

\begin{itemize}
\item[i)] $\widetilde{\mathcal{R}}(X,Y,Z,W)=-\widetilde{\mathcal{R}}(Y,X,Z,W)$;

\item[ii)] $\widetilde{\mathcal{R}}(X,Y,Z,W)+\widetilde{\mathcal{R}%
}(Y,Z,X,W)+\widetilde{\mathcal{R}}(Z,X,Y,W)=0$;

\item[iii)] $\widetilde{\mathcal{R}}(X,Y,Z,W)=-\widetilde{\mathcal{R}%
}(X,Y,W,Z)$;

\item[iv)] $\widetilde{\mathcal{R}}(X,Y,Z,W)=\widetilde{\mathcal{R}}(Z,W,X,Y)$.
\end{itemize}

If $\{\partial_{i}\}$ is a local basis for $T_{p}\mathcal{M}$, the curvature
tensor $\widetilde{\mathcal{R}}(X,Y,Z,W)$ can be calculated as follows
\begin{align}
&  \widetilde{\mathcal{R}}_{ijkm}=\widetilde{\mathcal{R}}(\partial
_{i},\partial_{j},\partial_{k},\partial_{m})=\\
&  \left(  \partial_{i}\Gamma_{jk}^{s}-\partial_{j}\Gamma_{ik}^{s}\right)
g_{sm}+\left(  \Gamma_{irm}\Gamma_{jk}^{r}-\Gamma_{jrm}\Gamma_{ik}^{r}\right)
\text{.}\nonumber
\end{align}
Furthermore, the sectional curvature $\mathcal{K}(\sigma_{X,Y})$ is given by,
\begin{equation}
\mathcal{K}(\sigma_{X,Y})\overset{\text{def}}{=}\frac{g(R(X,Y)Y,X)}%
{g(X,X)g(Y,Y)-g(X,Y)^{2}}\text{,} \label{SectionalCurv}%
\end{equation}
where $\sigma_{X,Y}$ is the $2$-plane section spanned by $X$ and $Y$. In a
Riemannian manifold, $\mathcal{K}(\sigma_{X,Y})$ completely determines the
curvature tensor. The converse is also true whenever $\mathcal{R}$ satisfies
properties i)-iv). Two other contractions of the curvature tensor are of
interest in our manuscript, the Ricci curvature and the scalar curvature. In
local coordinates, the \emph{Ricci curvature }can be written as,
\begin{equation}
\mathcal{R}\emph{ic}\overset{\text{def}}{=}\mathcal{R}_{ij}dx^{i}\otimes
dx^{j}\text{, with }\mathcal{R}_{ij}=\widetilde{\mathcal{R}}_{kij}^{k}\text{.}%
\end{equation}
The \emph{scalar curvature} is the function $S=S(p)$ defined as the trace of
the Ricci curvature $\mathcal{R}\emph{ic}$,%
\begin{equation}
S(p)\overset{\text{def}}{=}\text{\textrm{tr}}_{g}\left(  \mathcal{R}%
\emph{ic}\right)  =\mathcal{R}_{i}^{i}={\displaystyle\sum\limits_{i,j\text{
with }i\neq j}}\mathcal{K}\left(  u_{i}\text{, }u_{j}\right)  \text{,}%
\end{equation}
where $\left\{  u_{k}\right\}  $ with $1\leq k\leq n$ is an orthonormal system
in $T_{p}\mathcal{M}$.

\medskip

An important geometric aspect of the Riemannian curvature tensor in Eq.
\eqref{RCurv} is its connection to the behavior of the geodesics obtained by
integrating the nonlinear coupled ordinary differential equations in Eq.
\eqref{geodesic}. In particular, the Riemannian curvature tensor characterizes
the geodesic deviation equation that describes the deviation of two geodesics
which, although assumed initially parallel to each other, may depart from each
other in the presence of a gravitational field. The way for measuring the
geodesic deviation relies on the definition of the Jacobi vector (or geodesic
separation) field and the description of how it changes along a geodesic
\cite{P07}. Let us proceed first to define the geodesic separation field and
then derive its evolution equation along a given geodesic.

Consider a geodesic $\gamma:[0,1]\rightarrow\mathcal{M}$ with respect to a
connection $\nabla$. A Jacobi variation of $\gamma$ is a smooth map
$\Sigma:(-\varepsilon,\varepsilon)\times\lbrack0,1]\rightarrow\mathcal{M}$
such that $\Sigma(0,t)\equiv\gamma(t)$ and for any $s\in(-\varepsilon
,\varepsilon)$ the curve
\begin{equation}
\Sigma_{s}(t)=\Sigma(s,t),\quad t\in\lbrack0,1] \label{geodvar1}%
\end{equation}
is a geodesic. In addition, for every fixed $t\in\lbrack0,1]$, the Jacobi
variation defines a smooth curve
\begin{equation}
\Sigma^{t}(s)=\Sigma(s,t),\quad s\in(-\varepsilon,\varepsilon)
\end{equation}
and a vector field
\begin{equation}
V(s)=\frac{\partial\Sigma^{t}}{\partial s}(s),\quad s\in(-\varepsilon
,\varepsilon).
\end{equation}
The Jacobi vector field on the geodesic $\gamma(t)$ is then defined as
\begin{equation}
J(t)\overset{\text{def}}{=}\frac{\partial\Sigma}{\partial s}(0,t),\quad
t\in\lbrack0,1]. \label{Jvector}%
\end{equation}
On the contrary, given two vectors $X,Y\in T_{p}M$, there exists on the
geodesic $\gamma(t)$ a Jacobi field $J(t)$ such that
\begin{equation}
J(0)=X,\qquad\frac{\nabla J}{dt}(0)=Y,
\end{equation}
where $\frac{\nabla J}{dt}$ is the covariant derivative of $J$ that reads in
local coordinates as follows,
\begin{equation}
\frac{\nabla J^{k}}{dt}=\frac{dJ^{k}}{dt}+\Gamma_{ij}^{k}\dot{\gamma}^{i}%
J^{j}.
\end{equation}
This vector $J(t)$ can be constructed by considering a curve $\alpha(s)$ such
that
\begin{equation}
\alpha(0)=\gamma(0),\qquad\dot{\alpha}(0)=X,
\end{equation}
and by constructing on $\alpha(s)$ a vector field $A(s)$ such
\begin{equation}
A(0)=\dot{\gamma}(0),\qquad\frac{\nabla A}{dt}(0)=Y.
\end{equation}
At this point we can characterize the Jacobi field $J(t)$ as the solution of
the Jacobi Levi-Civita equation \cite{P07},
\begin{equation}
\frac{\nabla^{2}J}{dt^{2}}(t)+\mathcal{R}\left(  J(t),\dot{\gamma}(t)\right)
\dot{\gamma}(t)=0, \label{JLCEq}%
\end{equation}
that in local coordinates reads
\begin{equation}
\frac{\nabla^{2}J^{k}}{dt^{2}}+\mathcal{R}_{ijr}^{k}\frac{d\gamma^{i}}%
{dt}J^{j}\frac{d\gamma^{r}}{dt}=0.
\end{equation}
Let us now briefly comment on the stability of geodesics. Consider again a
geodesic $\gamma(t)$ with respect to the connection $\nabla$. Consider also a
perturbed geodesic as
\begin{equation}
\widetilde{\gamma}^{i}(t)=\gamma^{i}(t)+J^{i}(t)
\end{equation}
where $J(t)$ is a Jacobi field. Therefore from Eq. \eqref{JLCEq} it is clear
that the stability of geodesics is completely encoded in the curvature tensor
$\mathcal{R}$. In the particular case of constant sectional curvature
$\mathcal{K}$ of the Eq. \eqref{SectionalCurv} we can clearly see how the
geodesic deviation works. Firstly, Eq. \eqref{JLCEq} becomes
\begin{equation}
\frac{\nabla^{2}J^{k}}{dt^{2}}+\mathcal{K}\ J^{k}=0, \label{JLCKconst}%
\end{equation}
where $\mathcal{K}$ is the constant sectional curvature of the manifold.
Choosing a geodesic frame, covariant derivatives become ordinary derivatives,
so that the solution of \eqref{JLCKconst}, with initial conditions $J(0)=0$,
$dJ(0)/dt=w(0)$, is
\begin{equation}
J(t)=\left\{
\begin{array}
[c]{ll}%
\frac{w(0)}{\sqrt{\mathcal{K}}}\sin\left(  \sqrt{\mathcal{K}}\ t\right)  &
(\mathcal{K}>0);\\
& \\
t\ w(0) & (\mathcal{K}=0);\\
& \\
\frac{w(0)}{\sqrt{-\mathcal{K}}}\sinh\left(  \sqrt{-\mathcal{K}}\ t\right)  &
(\mathcal{K}<0).
\end{array}
\right.
\end{equation}
It is evident that, in the particular case of an isotropic manifold (sectional
curvature $\mathcal{K}$ constant), geodesics are unstable if $\mathcal{K}<0$.

\pagebreak

\end{document}